\documentclass[conference]{IEEEtran}
\IEEEoverridecommandlockouts

\usepackage{cite}
\usepackage{amsmath,amssymb,amsfonts}
\usepackage{algorithmic}
\usepackage{graphicx}
\usepackage{textcomp}
\usepackage{changepage}
\usepackage{tabularx}
\usepackage{comment}  
\usepackage{algorithm}
\usepackage{subcaption}
\usepackage{lipsum}
\usepackage{caption}
\usepackage{xspace}

\usepackage[letterpaper, top=.75in, bottom=1in, left=0.625in, right=0.625in]{geometry}
\usepackage{xcolor}
 \def\BibTeX{{\rm B\kern-.05em{\sc i\kern-.025em b}\kern-.08em
    T\kern-.1667em\lower.7ex\hbox{E}\kern-.125emX}}

\begin{document}
\title{Lightweight Deep Learning-Based Channel Estimation for RIS-Aided Extremely Large-Scale MIMO Systems on Resource-Limited Edge Devices}
\author{Muhammad Kamran Saeed, Ashfaq Khokhar,~\IEEEmembership{Fellow,~IEEE}, and Shakil Ahmed,~\IEEEmembership{Member,~IEEE}
\vspace*{-0.80 cm}
\thanks{This research was partly funded by Palmer Department Chair Endowment Funds at Iowa State University.}
\thanks{ M. K. Saeed is with the Dept. of Electrical and Computer Engineering, Iowa State University, Ames, IA, USA (email: muhkamransaeed@gmail.com). A. Khokhar is with the College of Engineering, Kansas State University, Manhattan, KS, USA (email: akhokhar@k-state.edu). S. Ahmed is with the Dept. of Computer Science, College of Computing, Grand Valley State University, Allendale, MI, USA (email: ahmeshak@gvsu.edu).}

}
\mark{IEEE COMPUTING AND NETWORKING}
{}

\maketitle

\begin{abstract}

Next-generation wireless technologies such as Sixth Generation (6G) aim to meet demanding requirements such as ultra-high data rates, low latency, and enhanced connectivity. Extremely Large-Scale MIMO (XL-MIMO) and Reconfigurable Intelligent Surface (RIS) are key enablers, with XL-MIMO boosting spectral and energy efficiency through numerous antennas, and RIS offering dynamic control over the wireless environment via passive reflective elements. However, realizing their full potential depends on accurate Channel State Information (CSI). Recent advances in deep learning have facilitated efficient cascaded channel estimation. However, the scalability and practical deployment of existing estimation models in XL-MIMO systems remain limited. The growing number of antennas and RIS elements introduces a significant barrier to real-time and efficient channel estimation, drastically increasing data volume, escalating computational complexity, requiring advanced hardware, and resulting in substantial energy consumption. To address these challenges, we propose a lightweight deep learning framework for efficient cascaded channel estimation in XL-MIMO systems, designed to minimize computational complexity and make it suitable for deployment on resource-constrained edge devices. Using spatial correlations in the channel, we introduce a patch-based training mechanism that reduces the dimensionality of input to patch-level representations while preserving essential information, allowing scalable training for large-scale systems. Simulation results under diverse conditions demonstrate that our framework significantly improves estimation accuracy and reduces computational complexity, thereby maintaining robust performance even with increasing system dimensions in XL-MIMO networks.
\end{abstract}

\begin{IEEEkeywords}
Reconfigurable Intelligent Surface, B6G, Deep Learning, Channel Estimation, Extremely Large-Scale MIMO, Spatial Correlation in Wireless Channels.
\end{IEEEkeywords}

\section{Introduction}
Next-generation communication technologies, such as Sixth Generation (6G) and beyond, are expected to meet the rapidly growing demand for advanced service requirements \cite{RIS6G}. These include ultra-high data rates, extremely low latency, minimal bit error rates, enhanced device connectivity, and much more. To address these challenges, Extremely Large-Scale MIMO (XL-MIMO) has emerged as a transformative paradigm, offering significantly higher spectral and energy efficiency while ensuring reliable communication \cite{XL1}. Unlike conventional MIMO systems, XL-MIMO integrates thousands of antennas in a compact space, fundamentally altering electromagnetic characteristics \cite{XL2}. This not only expands the number of antennas, but also enhances spectral efficiency and spatial degrees of freedom, enabling unprecedented improvements in wireless communication performance \cite{XL3}. Building on existing advantages, the integration of Reconfigurable Intelligent Surfaces (RIS) with XL-MIMO presents a promising approach to address the growing demands of 6G and beyond networks \cite{RIS6G}. RIS technology is emerging as a transformative enabler, offering the ability to dynamically shape wireless channels between users and the Base Station (BS), thereby enhancing the overall performance of the system \cite{RIS}. RIS panels consist of tunable electromagnetic elements, each independently controlled to direct incident waves towards targeted directions \cite{OP1,BLMMSE}. This adaptability optimizes the propagation environment, significantly boosting the spectral efficiency \cite{RIS1}. Furthermore, RIS elements are passive, energy-efficient, and easily integrated into surfaces, offering a low-cost solution to enhance signal quality and network coverage \cite{OP2}.

Realizing the performance gains of RIS-enabled XL-MIMO systems in practice requires optimizing the design parameters, which are highly dependent on accurate Channel State Information (CSI) \cite{OP3}. However, obtaining CSI in RIS is challenging because RIS elements are passive and do not receive or transmit signals directly \cite{DL1}. Since individual channels between user–RIS and RIS–BS cannot be estimated separately, only the cascaded channel can be inferred, which encompasses the link from the user to the BS via the RIS. This limitation presents distinct challenges in accurately estimating the overall channel. Firstly, unlike traditional Rayleigh fading channels, RIS-based channels deviate from conventional models, resulting in a performance gap between Minimum Mean Square Error (MMSE) and Linear-MMSE (LMMSE) estimators \cite{RIS6G}. Moreover, the large number of elements in RIS and antenna arrays complicates cascaded channel estimation, significantly expanding the parameter space and making it resource-intensive for edge devices~\cite{DL2, OP1}. Consequently, these challenges limit system performance and hinder the large-scale deployment of RIS-assisted XL-MIMO systems. This paper proposes a power-efficient data-driven approach for cascaded channel estimation at the edge devices for RIS-enabled XL-MIMO systems.

Recent work in cascaded channel estimation can be broadly categorized into two main sets of approaches: conventional model-driven methods \cite{OP1, OP2, OP3} and data-driven techniques \cite{DL1, DL2, DL3, DL4, DL5, DL6}. Conventional model-driven methods frame cascaded channel estimation as an optimization problem, solved iteratively to improve accuracy. Although these methods have achieved notable advancements, the need for multiple iterative steps in real-time scenarios makes them time-consuming, which is impractical for future low-latency communications \cite{DL1}. In contrast, data-driven approaches (such as machine and deep learning) are gaining popularity. These techniques offer minimal inference time, making them well suited for real-world, time-sensitive scenarios. Although data-driven methods require relevant training data, they can significantly enhance the efficiency and accuracy of channel estimation once trained~\cite{DL5}.

Recent advances in deep learning have substantially contributed to RIS-based communication systems, delivering robust performance across diverse environmental conditions and enabling low-complexity inference. Various deep learning approaches have been proposed to improve cascaded channel estimation for massive MIMO systems, significantly improving overall system performance~\cite{DL5, DL1, DL2, DL3, DL4, DL6}.  However, these techniques face scalability challenges and are typically limited to massive MIMO systems, where only tens of antennas and BS elements are considered. Extending the training setup to RIS-enabled XL-MIMO systems, involving hundreds of RIS elements and thousands of antennas as proposed in these models, would not be a viable solution due to various reasons. Firstly, these techniques use tens of thousands of samples to train their models. Processing and storing such large datasets in RIS-assisted XL-MIMO-based systems can lead to inefficiencies, introduce substantial computational and memory demands, and increase both training time and resource consumption~\cite{DL1}. Furthermore, training these models on large datasets requires substantial energy resources, making them not only energy inefficient but also raising environmental concerns due to increasing energy consumption. Additionally, the need for highly specialized high-performance hardware to handle the training of such expensive systems adds considerable costs, making it financially challenging to scale these solutions. Finally, with edge computing gaining attention in recent years as a key enabler for distributed low-latency applications, the deployment of these complex models in edge environments becomes a significant obstacle \cite{FL}. The limited computational resources and energy constraints at the edge create a major barrier, hindering the practical implementation of these techniques in future communication networks.

Addressing the aforementioned challenges requires a compact and efficient representation of the cascaded channel. We, therefore, exploit the spatial correlation present in the cascaded channel, as discussed in \cite{sp1,sp2,sp3}, to enable computationally efficient model training suitable for edge devices in resource-limited settings. To take advantage of this property,  we introduce a patch-based training framework that leverages the inherent spatial correlations in the cascaded channel. In the proposed framework, a few small, localized patches are extracted from each high-dimensional channel to effectively capture salient features necessary for accurate channel estimation. This patch-wise processing significantly reduces input dimensionality and overall system complexity, thereby enabling scalable and computationally efficient training, particularly suitable for XL-MIMO systems deployed at the edge. Furthermore, employing efficient deep learning techniques, such as encoder-decoder architecture, enables the transformation of representative patches into compact latent representations, thereby facilitating more effective learning of spatial representation~\cite{UNET1}. These latent features effectively preserve and highlight spatial dependencies between RIS elements and BS antennas, facilitating the learning of inherent correlations while substantially reducing training complexity. Furthermore, leveraging the encoder-decoder architecture also helps break the correlation of noise in the data, effectively isolating meaningful patterns from unwanted distortions \cite{AE1}. For example, in \cite{GM2}, the model demonstrated strong potential for channel estimation in MIMO systems.

Motivated by these advantages, our proposed channel estimation framework adopts an encoder–decoder architecture with multi-scale feature extraction and aggregation, enabling more effective representation of salient structures in noisy inputs and, in turn, more accurate channel estimation. This design captures spatial features efficiently, particularly benefiting resource-constrained environments such as edge devices. We propose a deep learning-based solution for channel estimation in RIS-assisted XL-MIMO systems, formulated as a denoising problem that recovers clean channel information from noisy observations. The proposed method improves estimation accuracy and reduces computational complexity, remaining robust as the number of antennas and RIS elements grows in XL-MIMO systems. Extensive simulations under diverse conditions validate the scalability and effectiveness of the proposed framework, demonstrating substantial gains in cascaded channel estimation and system efficiency for RIS-enabled XL-MIMO deployments.

\subsection{Literature Review}
Conventional model-driven methods remain widely used because they operate without training data and can be deployed directly. In~\cite{OP1}, sparsity and structural correlation in mmWave channels are exploited to jointly optimize the RIS phase shifts and minimize the pilot overhead required for successive channel updates. A binary reflection scheme in~\cite{BLMMSE} activates one RIS element at a time while keeping the others off, obtaining channel estimates via least squares (LS). In \cite{lee2025nearfield}, the authors model the BS--RIS channel using geometric free-space propagation and array responses. They partition the effective channel into piecewise low-rank segments and jointly recover them using collaborative low-rank approximation, reducing channel estimation complexity. In~\cite{bian2026sparse}, a three-stage signal-assisted sparse representation scheme combines double-domain angle/polar-domain filtering, failure-aware orthogonal matching pursuit, and double-sparsity refinement to jointly recover the channel and diagnose RIS element failures under low pilot overhead. In~\cite{zhong2025hybrid}, a hybrid-field stochastic gradient pursuit algorithm addresses hybrid-field XL-MIMO by jointly exploiting angular-domain sparsity for far-field users and polar-domain sparsity for near-field users, with a least mean squares step for reduced complexity. Other formulations include an alternating optimization with a covariance method solved via successive convex approximation~\cite{OP2}, and a compressive sensing framework with a two-step joint estimator that exploits block-sparsity shared across multiple users~\cite{OP3}. Despite their adequate performance, conventional model-driven methods often rely on iterative schemes that scale poorly with hundreds of RIS elements and thousands of antennas, making them impractical for real-time XL-MIMO deployment.

The pioneering work that introduced deep learning for cascaded channel estimation was presented in~\cite{DL5}. The proposed framework consisted of two convolutional neural networks (CNNs), each comprising nine layers, designed to estimate both the direct channel between the BS and the user and the cascaded channel between the BS and the user through an RIS. Moreover, in~\cite{DL1}, the authors proposed a deep residual network (CDRN) method using a CNN-based architecture to effectively remove residual noise from noisy pilot signals, ultimately enhancing estimation accuracy. Additionally,~\cite{DL2} introduced a bidirectional long short-term memory (Bi-LSTM) approach for channel estimation. Bi-LSTM processed past and future information simultaneously through two LSTM layers, thereby enhancing feature extraction and achieving adequate accuracy in channel estimation. In~\cite{zheng2024convolutional}, hybrid-field channel estimation for XL-RIS-aided massive MIMO was formulated as a convolutional dictionary learning problem cast into a bilevel optimization framework, solved via an unrolled proximal gradient descent-based network.

Similarly, the work in \cite{DL3} interprets the received pilot signal as a low-resolution image and formulates channel estimation as a super-resolution reconstruction problem. This approach employed a CNN to extract detailed features, followed by a denoising CNN to eliminate additive noise. Another study in \cite{DL6} proposed a model-driven deep unfolding approach for cascaded channel estimation. The proposed scheme leverages the rank deficiency of the cascaded channel to achieve accurate estimation while significantly reducing communication overhead. In~\cite{DL4}, a self-supervised learning framework was developed for channel estimation, designed to operate under both supervised and unsupervised conditions. Specifically, in the absence of ground-truth data, the method synthesized input samples by introducing controlled noise into the received signal, using the resulting noisy observations as the corresponding labels. However, in real-world scenarios, accurate estimates of the noise variance are often unavailable. Although the techniques mentioned above~\cite{DL5, DL1, DL2, DL3, DL4, DL6} demonstrate satisfactory performance, they still encounter challenges related to scalability and estimation accuracy and rely on unrealistic dataset assumptions that may not adequately reflect real-world scenarios. 

A significant limitation of data-driven techniques, as mentioned earlier, is that most proposed schemes are trained and evaluated on small-scale systems, typically involving only tens of RIS and antenna elements. Although these schemes perform effectively in moderate-scale scenarios, their scalability becomes a critical bottleneck when applied to large-scale systems comprising hundreds of RIS elements and thousands of antennas, rendering them impractical for RIS-enabled XL-MIMO deployments~\cite{RIS}. Training such large systems requires complex, high-end computational infrastructure to meet the associated processing demands, posing a major challenge to the feasibility of these approaches. Moreover, several studies~\cite{RIS, OP1, OP3} have highlighted the presence of spatial correlation in the cascaded channel. In particular, some works have conducted detailed analyses of spatially correlated RIS and antenna elements, providing valuable insights into the impact of correlation on system performance~\cite{sp1,sp2,sp3}. Motivated by these observations, this paper leverages the spatial correlation inherent in the cascaded channel to reduce model complexity, thereby enabling efficient deployment of XL-MIMO systems on resource-constrained devices.

\subsection{Contributions}
We employ deep learning to address the challenges of cascaded channel estimation in RIS-assisted XL-MIMO systems. Specifically, the key contributions of this work are as follows:

\begin{itemize}

     \item We propose a lightweight deep learning framework for efficient cascaded channel estimation in RIS-enabled XL-MIMO systems. The proposed framework is designed with reduced computational complexity, rendering it highly suitable for deployment on resource-constrained edge devices, which are integral to the infrastructure of next-generation wireless communication networks.
    
    \item We leverage the spatial correlations in the channel to develop a patch-based training mechanism, in which a small number of carefully selected patches capture the essential features of each channel. This approach effectively reduces the input dimensionality to patch-level representations while preserving essential information, thereby facilitating scalable training for large-scale systems.

     \item We propose a multi-scale feature extraction and aggregation-based architecture, customized with a dedicated denoising module to effectively extract noise from the channel coefficients and enable efficient cascaded channel estimation. The proposed architecture demonstrates reliable performance even when trained using limited data, maintaining high accuracy with small-sized patches. This highlights its robustness, computational efficiency, and adaptability for deployment on low-resource edge devices.
     
    \item We generate a synthetic dataset based on realistic real-world settings to accurately capture the spatial characteristics of wireless communication environments, moving beyond purely random channel modeling for improved fidelity and reliability.
    
    \item Extensive simulations are conducted under various channel conditions, and the results show that the proposed solution maintains consistent complexity regardless of the size of the RIS-enabled XL-MIMO system, demonstrating its scalability and suitability for edge device deployment.

\end{itemize}

The remainder of the paper is organized as follows. Section~II introduces the system model and the proposed channel estimation framework. Section~III provides a detailed description of the proposed model architecture. Section~IV discusses the spatial correlation between antennas and RIS elements. Section~V presents the patch-based channel estimation approach along with an in-depth effectiveness analysis.  Section~VI analyzes the computational complexity of the proposed method. Section~VII  outlines the simulation settings and presents the results. Finally, Section~VIII concludes the paper with key findings.

\textit{Notation:} \( T \) and \( H \) represent the transpose and Hermitian transpose of a matrix, while \( \mathbb{R} \) and \( \mathbb{C} \) denote the sets of real and complex numbers, respectively. The function \( \text{diag}(\cdot) \) constructs a diagonal matrix, and \( \| \cdot \|_F \) denotes the Frobenius norm, and  \( e \) indicates the exponential function. \({\mathbf{I}_M}\) represents the identity matrix of size $M \times M$. The operator $(\cdot)^{-1}$ denotes the matrix inverse, while $\mathbb{E}(\cdot)$ indicates the statistical expectation. The symbol $\otimes$ is used to represent the Kronecker product. Lowercase bold letters (e.g., $\mathbf{y}$) represent vectors, uppercase bold letters (e.g., $\mathbf{Y}$) represent matrices, and non-bold letters denote scalars.

\section{System Model}
This paper considers an XL-MIMO system assisted by an RIS panel, for example deployed on the facade of a building, as shown in Fig.~1, to improve the propagation environment. The proposed system model includes a BS equipped with \( M \) antennas arranged in a uniform planar array configuration, where \( M_h \) and \( M_v \) are the numbers of horizontal and vertical antennas, communicating with \(K\) single-antenna users. Moreover, the RIS, with passive reflecting elements \( N = N_h \times N_v \), is arranged in a uniform planar array, where \( N_h \) and \( N_v \) are the numbers of horizontal and vertical elements, respectively. Additionally, each RIS element acts as a single point source, aggregating incoming signals and retransmitting them to the BS. We assume that the RIS and users lie in the far field of the BS array, and that the users likewise lie in the far field of the RIS. Specifically, for an array with largest aperture dimension $D$ and carrier wavelength $\lambda$, the Rayleigh (Fraunhofer) distance is given by $d_R = 2D^2/\lambda$. The BS--RIS, BS--user, and RIS--user distances are assumed to exceed $d_R$ for the corresponding aperture.

\begin{figure}[t]
    \centering
    \vspace{-3mm}
    \includegraphics[width=9cm]{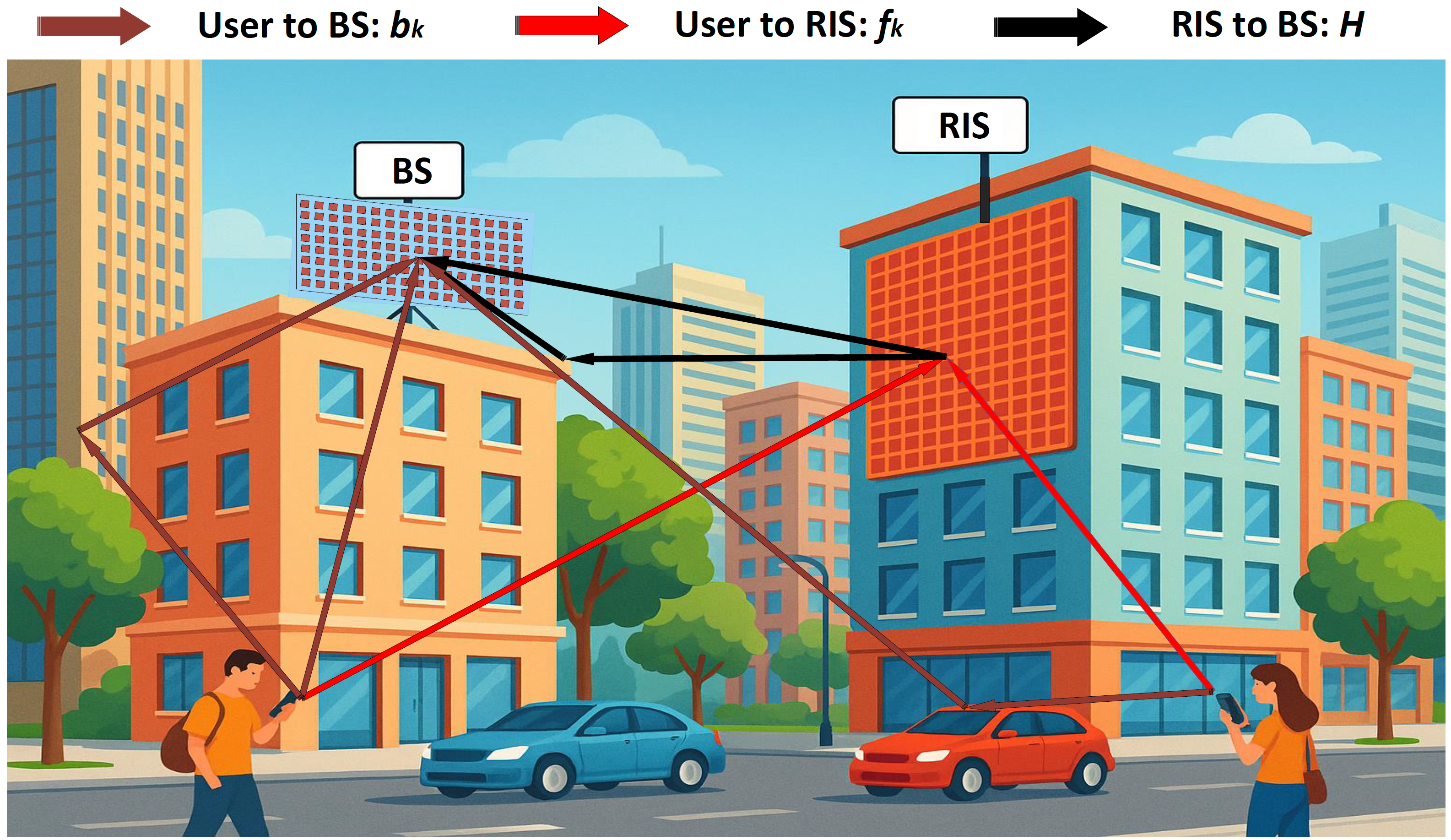}
    \caption{The uplink of the RIS-assisted XL-MIMO system.}
    \label{fig:1}
    \vspace{-5mm}
\end{figure}

We consider a real-world scenario where a signal propagates through a three-dimensional (3D) space, with a local spherical coordinate system defined such that \( \phi \) represents the azimuth angle and \( \psi \) denotes the elevation angle \cite{sp1}. The reflecting and antenna elements are arranged in a two-dimensional grid within the YZ plane. Each element has \( d_H \) and \( d_V \) corresponding to the horizontal and vertical spacing, respectively. To systematically index the elements, they are ordered row by row and denoted as \(n\in [1, N] \) RIS elements and \(m\in [1, M] \) antenna elements. The \(n\)-th element’s position relative to the origin is:

\begin{equation}
\boldsymbol{d}_n = \begin{bmatrix} 0, \alpha_n d_H, \gamma_n d_V \end{bmatrix}^T,
\end{equation}
where the horizontal \(\alpha_n\) and vertical \(\gamma_n\) indices of the \( n \)-th element are represented as:
\begin{equation}
\alpha_n = \text{mod}(n - 1, N_h), \quad \gamma_n = \lfloor (n - 1) / N_h \rfloor,
\end{equation}
where \( \text{mod}(\cdot, \cdot) \) denotes the modulus operation, and \( \lfloor \cdot \rfloor \) represents the floor function. Now, consider a plane wave with wavelength \( \lambda \) impinging on the RIS from an azimuth angle \( \phi \) and an elevation angle \( \psi \). The array response vector corresponding to this incident wave is given by:
\begin{equation}
\mathbf{a}(\phi, \psi) = \begin{bmatrix} e^{j \boldsymbol{\omega}(\phi, \psi)^T \boldsymbol{d}_1}, \dots, e^{j \boldsymbol{\omega}(\phi, \psi)^T \boldsymbol{d}_N} \end{bmatrix}^T,
\end{equation}
where \( \boldsymbol{\omega}(\phi, \psi) \in \mathbb{R}^{3 \times 1} \) denotes the wave vector, defined as:
\begin{equation}
\boldsymbol{\omega}(\phi, \psi) = \frac{2\pi}{\lambda} \begin{bmatrix} \cos(\psi) \cos(\phi) \\ \cos(\psi) \sin(\phi) \\ \sin(\psi) \end{bmatrix}.
\end{equation}

Furthermore, a Line of Sight (LoS) is assumed to exist between the users and the RIS and between the RIS and the BS. We define the channel between the \( k \)-th user and the RIS elements as a Rician fading channel represented by \( \mathbf{f}_k \in \mathbb{C}^{ N\times 1} \), which can be expressed as follows:
\begin{equation}
\mathbf{f}_k  = \sqrt{\frac{\eta_r}{\eta_r+1}} \mathbf{f}^{{LoS}}_k + \sqrt{\frac{1}{\eta_r+1}} \mathbf{f}^{{NLoS}}_k.
\end{equation}

Moreover, the LoS channel component \(\mathbf{f}^{LoS}_k\) between the user and RIS is given by:
\begin{equation}
\mathbf{f}^{LoS}_k =  \mathbf{a}_r^H(\phi_{r,k}, \psi_{r,k}),
\end{equation}
where \( \phi_{r,k} \) and \( \psi_{r,k} \) represent the azimuth and elevation angles of the \( k \)-th user's signal at the RIS. For a uniform planar array with \( N = N_h \times N_v \) reflecting elements, the array response vector \( \mathbf{a}_r(\phi, \psi) \in \mathbb{C}^{ N\times 1}\) can be modeled as:

\begin{equation}
\begin{split}
\!\!\mathbf{a}_r(\phi, \psi) \!\!= \!\!\frac{1}{\sqrt{N}} 
[1, \ldots,e^{j \frac{2\pi}{\lambda} \left( \alpha_n\cos(\psi) \sin(\phi)d_H + \gamma_n\sin(\phi)d_V \right)},  \\ 
\ldots,  e^{j \frac{2\pi}{\lambda}\left( \alpha_{N} \cos(\psi) \sin(\phi)d_H + \gamma_{N} \sin(\phi)d_V \right)  }]^T,
\end{split}
\end{equation}
where $\alpha_n$ and $\gamma_n$ are the indices of reflecting elements in the horizontal and vertical directions, respectively. Since the user--RIS distance is assumed to exceed the Rayleigh distance $d_R$ associated with the RIS aperture, the wavefront incident on the RIS from the user is well approximated as a plane wave, consistent with the far-field array-response model above. We consider the Non-LoS (NLoS) channels as quasi-static Rayleigh fading, which remains constant for the coherence interval. The NLoS channel between the RIS and the user is modeled using correlated Rayleigh fading \cite{sp0}. The NLoS channel from the RIS to the user is given by:
\begin{equation}
\mathbf{f}^{NLoS}_k = {(\mathbf{R}_{r})^\frac{1}{2}}  \Tilde{\mathbf{f_k}},
\end{equation}
where $\tilde{\mathbf{f}}_k \in \mathbb{C}^{N\times1}$ follows a complex Gaussian distribution $\mathcal{N}_{\mathbf{c}}(0,\mathbf{I}_N)$, and $\mathbf{R}_r$ denotes the RIS correlation matrix discussed in Section~III. The Rician RIS--BS channel, $\mathbf{H}\in\mathbb{C}^{M\times N}$, is then expressed as follows:

\begin{equation}
\mathbf{H} = \sqrt{\frac{\eta_b}{\eta_b+1}} \mathbf{H}^{{LoS}} + \sqrt{\frac{1}{\eta_b+1}} \mathbf{H}^{{NLoS}},
\end{equation}
where \( \mathbf{H}^{\text{LoS}} \) is the Kronecker product of the array response of the BS and the RIS, which can be expressed as follows:

\begin{equation}
\mathbf{H}^{LoS} =  \mathbf{a}_b (\phi_{b}, \psi_{b})\otimes\mathbf{a}_r^H(\phi_{r}, \psi_{r}) ,
\end{equation}
where \(\mathbf{a}_b (\phi_{b}, \psi_{b}) \in \mathbb{C}^{M \times 1}\)  and \(\mathbf{a}_r(\phi_{r}, \psi_{r}) \in \mathbb{C}^{N\times 1}\) are the array response vectors associated with the RIS and BS, respectively. Here, \(\phi_r\) and \(\psi_r\) denote the azimuth and zenith angles of departure at the RIS, while \(\phi_b\) and \(\psi_b\) denote the corresponding angles of arrival at the BS.  The array response vector for a uniform planar array at the BS can be expressed similarly to equation (7) as:

\begin{equation}
\begin{split}
\mathbf{a}_b(\phi, \psi) = \frac{1}{\sqrt{M}} 
[1, \ldots,e^{j \frac{2\pi}{\lambda} \left( \alpha_m\cos(\psi) \sin(\phi)d_{H} + \gamma_m\sin(\phi)d_{V} \right)},  \\ 
\ldots,  e^{j \frac{2\pi}{\lambda}\left( \alpha_{M} \cos(\psi) \sin(\phi)d_{H} + \gamma_{M} \sin(\phi)d_{V} \right)  }]^T,
\end{split}
\end{equation}
where \( \alpha_m \) and \( \gamma_m \) are the indices of antenna elements in the horizontal and vertical directions, respectively. The \(\mathbf{H}^{NLoS}\) can be expressed as:

\begin{equation}
\mathbf{H}^{{ NLoS}} = {(\mathbf{R}_{b})^\frac{1}{2}}  \Tilde{\mathbf{H}} {(\mathbf{R}_{r})^\frac{1}{2}},
\end{equation}
where \( \Tilde{\mathbf{H}}\) \(\sim\) \( \mathcal{N_\textbf{c}}(0, \mathbf{I}_M \otimes \mathbf{I}_N)  \) and \( \mathbf{R}_{\text{b}} \) and \( \mathbf{R}_{\text{r}} \) are the correlation matrices for the BS and RIS, respectively. The correlation matrix for the BS is modeled as a Toeplitz matrix using an exponential correlation function \( \mathbf{R_b}_{[i,j]} = \rho^{|i-j|} \) between \(i \)-th and \(j \)-th antenna, where \( \rho \) is the correlation factor, which characterizes the correlation between elements based on their relative indices. This exponential model assumes that the correlation decays with the distance between elements. The total 2D correlation matrix \( \mathbf{R}_b \) is then obtained by taking the Kronecker product of the horizontal and vertical correlation matrices, i.e., \( \mathbf{R}_b = \mathbf{R}_{M_h} \otimes \mathbf{R}_{M_v} \), where \( \mathbf{R}_{M_h} \) and \( \mathbf{R}_{M_v} \) are the correlation matrices for the horizontal and vertical directions, respectively. Moreover, the direct channel between the  \( k \)-th user and the BS follows a correlated Rayleigh fading model represented by \( \mathbf{b}_k \in \mathbb{C}^{ M\times 1} \), expressed as:

\begin{equation}  
\mathbf{b}_k = (\mathbf{R}_b)^{\frac{1}{2}} \Tilde{\mathbf{b_k}},  
\end{equation}  
where \( \Tilde{\mathbf{b_k}} \) follows a complex Gaussian distribution \( \mathcal{N_\textbf{c}}(0, \mathbf{I}_M) \), and \( \mathbf{R}_b \) represents the correlation matrix for the BS.

\subsection{Channel Estimation Framework }
The system's performance is heavily dependent on the availability of precise channel information. We adopted a Time Division Duplex (TDD) protocol, which exploits channel reciprocity to estimate downlink CSI from uplink CSI. The TDD protocol divides communication into time slots, known as coherence intervals, during which the channel response remains approximately constant \cite{sp0}. We divide the channel estimation framework into two phases: direct channel estimation and cascaded channel estimation. In the first phase, the RIS is turned off, so no reflections occur. The direct channel between the user and the BS is estimated during this stage, as shown in Fig. 2. Moreover, in the second phase, we estimate the cascaded channel between the user, the RIS, and the BS. Estimating the cascaded channel requires \( L \) distinct phase shift matrices, where \(L\geq N \), to facilitate accurate channel estimation. For each phase shift matrix, each user is assigned an orthogonal pilot sequence \( \mathbf{x}_k \) of length \( u \), where \( u \geq K \), ensuring the pilot's orthogonality for accurate estimation. For the \( k \)-th user during the \( l \)-th subframe, the pilot signal received at the BS, denoted as \( \mathbf{y}_k \in \mathbb{C}^{M \times 1} \), is given by the following expression:

\begin{figure}[t]
    \vspace{-3mm}

    \centering
    \includegraphics[width=8cm]{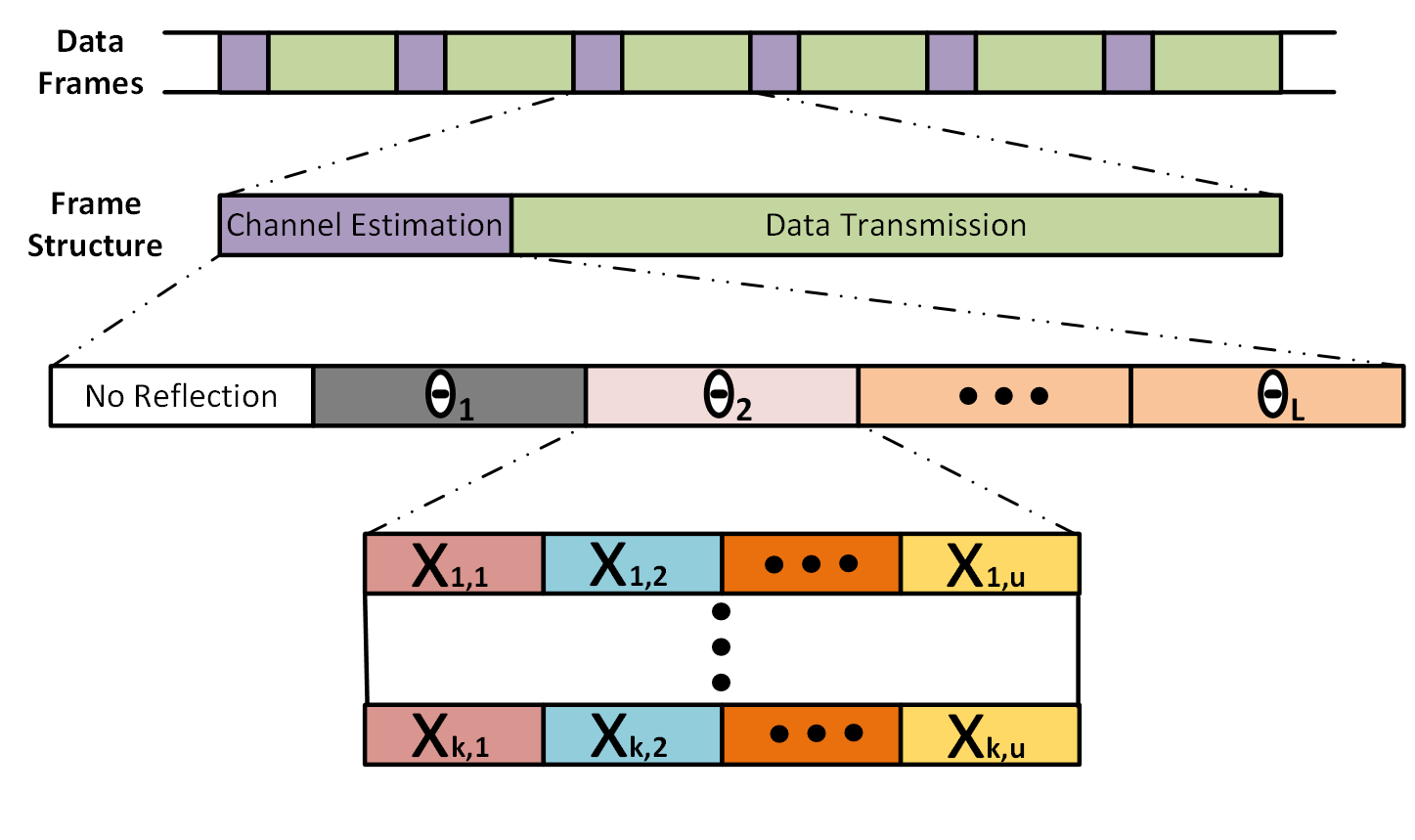}
    \caption{Channel estimation framework has two phases: (1) Direct channel estimation with the RIS off, and (2) Cascaded channel estimation with the RIS on using \( L \ge N\) distinct phase shift matrices. }

    \vspace{-3mm}

    \label{fig:3}
\end{figure}

\begin{equation}
\mathbf{y}_k = \mathbf{b}_k\mathbf{x}_k^H +\mathbf{H} \text{diag}(\mathbf{\Theta}_l) \mathbf{f}_k \mathbf{x}_k^H + \mathbf{n}_k,
\end{equation}
where \( \boldsymbol{\Theta}_l = \begin{bmatrix} \beta_{l,1} e^{j\theta_{l,1}}, \beta_{l,2} e^{j\theta_{l,2}}, \ldots, \beta_{l,N} e^{j\theta_{l,N}} \end{bmatrix}^T \) denotes the reflection coefficients of the RIS elements in the \( l \)-th sub-frame. Here, \( 0 \leq \beta_{l,n} \leq 1 \) and \( 0 \leq \theta_{l,n} < 2\pi \) indicate the amplitude and phase shift of the \( n \)-th RIS element, respectively.
 The noise term \( n_k \sim \mathcal{CN}(0, \sigma^2 \mathbf{I}_m) \) represents AWGN. Furthermore, we can thus separate the \(k\)-th user signal by multiplying the received signal with \( \frac{1}{Pu} x_k \):

\begin{equation}
\mathbf{y}_k =  \frac{1}{Pu}( \mathbf{b}_k \mathbf{x}_k^H  \mathbf{x}_k + \mathbf{H} \text{diag}(\mathbf{\Theta}_l) \mathbf{f}_k \mathbf{x}_k^H  \mathbf{x}_k + \mathbf{n}_k \mathbf{x}_k ).
\end{equation}

Since pilot sequences are orthogonal, \( \mathbf{x}_k^H\mathbf {x}_k = Pu \), where \( P \) and \( u \) are the power and length of the pilot, respectively. Here the noise term \( \mathbf{v_k} = \frac{1}{Pu} \mathbf{x}_k \mathbf{n}_k \):

\begin{equation}
\mathbf{y}_k = \mathbf{b}_k + \mathbf{H}  \text{\text{diag}} (\mathbf{\Theta}_l) \mathbf{f}_k  + \mathbf{v}_k
\end{equation}

It is evident that 
\(
\text{\text{diag}}(\mathbf{\Theta}_l) \mathbf{f}_k = \text{\text{diag}}(\mathbf{f}_k) \mathbf{\Theta}_l
\). For  cascaded channel estimation, the process requires received pilot signals across \( L \) frames, collected as 
\(
\mathbf{Y}_k = [\mathbf{y}_{k,1}, \mathbf{y}_{k,2}, \ldots, \mathbf{y}_{k,L}].
\)
Similarly, the phase shift matrices across these frames are represented by 
\(
\mathbf{S} = [ \mathbf{\Theta}_1, \mathbf{\Theta}_2, \ldots, \mathbf{\Theta}_L] \in \mathbb{C}^{N \times L}.
\)

\begin{equation}
\mathbf{Y}_k =\mathbf{b}_{k} + \mathbf{G}_k \mathbf{S}  + \mathbf{V}_k,
\end{equation}
where \( \mathbf{G}_k = \ \mathbf{H} \, \text{diag}(\mathbf{f}_k)  \in \mathcal{C}^{ M \times N} \) represents the cascaded channel and \(
\mathbf{V}_k = [\mathbf{v}_{k,1}, \mathbf{v}_{k,2}, \ldots, \mathbf{v}_{k,L}]
\). According to \cite{DL4}, designing the optimal phase shift matrix \( \mathbf{S} \) using a discrete Fourier transform maximizes the received signal power at the BS, which can be expressed as follows:
 
\begin{equation}
\mathbf{S} = 
\begin{bmatrix}

1& W^{1} &  \cdots & W^{1.(L-1)} \\
\vdots  & \vdots & \ddots & \vdots \\
1&W^{N} & \cdots & W^{N.(L-1)}
\end{bmatrix}
\in \mathbb{C}^{N \times L},
\end{equation}
where \( W = e^{j \frac{2\pi}{L}} \) and \( \mathbf{S} \mathbf{S}^H = L \mathbf{I}_{N} \).   Cascaded channel estimation involves recovering \( \mathbf{G_k} \) from the received pilots. The simplest approach is the LS method \cite{LS}, which performs effectively when the channel is assumed to be static. The LS representation of the cascaded channel can be expressed as \( \mathbf{\Tilde{G}}_k^{LS} = \mathbf{Y}_k\mathbf {S}^H (\mathbf{S} \mathbf{S}^H)^{-1} \), where \( \mathbf{\Tilde{G}}^{LS} = [ \mathbf{\Tilde{G}}_1^{LS},\mathbf{\Tilde{G}}_2^{LS},\ldots,\mathbf{\Tilde{G}}_\tau^{LS}] \), and where \( \tau \) represent total training data. However, treating the channel as static fails to capture its spatial characteristics, especially when compared to data-driven methods. Moreover, under a Rayleigh fading model and assuming the availability of the statistical channel correlation matrix, the optimal estimator is the MMSE estimator. However, when the channel distribution is complex, as in the case of the cascaded channel in our scenario, the MMSE estimator becomes computationally expensive. To reduce this complexity, the MMSE estimator is often approximated by the LMMSE estimator. According to \cite{LMMSE}, the LMMSE estimator is given by:

\begin{equation} 
\mathbf{\Tilde{G}}_k^{\mathrm{LMMSE}} = \mathbf{Y}_k (\mathbf{S}^H \mathbf{R}_{G_k} \mathbf{S} + M \sigma_V^2 \mathbf{I}_S)^{-1} \mathbf{S}^H \mathbf{R}_{G_k},
\end{equation}
where \( \mathbf{R}_{G_k} = \mathbb{E} (G_k^H G_k) \) denotes the statistical channel correlation matrix. However, since \( G_k \) corresponds to the \( k-th\) user's cascaded channel, it does not follow the Rayleigh fading model. Consequently, the assumptions underlying the LMMSE estimator are violated, leading to a significant performance gap between the LMMSE estimation and the actual channel. To address this limitation, we propose a deep learning-based method that effectively learns the underlying channel structure and mitigates the performance gap introduced by the model mismatch in conventional methods.

\section{Deep Learning Assisted Channel Estimation}
Deep learning-assisted cascaded channel estimation has gained significant popularity in recent years~\cite{DL5, DL1, DL2, DL3, DL4, DL6}. One of the key reasons for this growing interest is its extremely low inference time, which makes it well-suited for low-latency applications. However, existing works primarily focus on small-scale scenarios with only a limited number of antennas and RIS elements. In this paper, we investigate the use of deep learning to estimate the channel in RIS-enabled XL-MIMO systems, which may involve hundreds of RIS elements and thousands of antennas. We leverage the spatial correlation inherent in the cascaded channel, resulting from the closely placed RIS elements and BS antennas. Building on this property, we propose a patch-based training framework where small, localized patches are extracted from each high-dimensional data point \( \mathbf{\Tilde{G}}^{\mathrm{LS}}_i \in \mathbb{C}^{M \times N} \), capturing salient features while reducing input dimensionality and enabling scalable training for large-scale systems on resource-constrained edge devices. This section presents the proposed deep learning framework, while Sections IV and V discuss spatial correlation and the patch-based training strategy.

\subsection{Theoretical Background}
Recent studies have demonstrated that encoder–decoder architectures can effectively learn compact, low-dimensional representations by capturing essential patterns and structures while minimizing reconstruction loss~\cite{AE1, UNET1, GM2}. These architectures also exhibit strong generalization capabilities, making them suitable for modeling complex, high-dimensional data distributions. Moreover, they typically require less training data, as they learn latent representations that preserve the most critical features of the input. This latent space not only facilitates efficient data compression and generalization but also enables the extraction of meaningful features, which is particularly beneficial in our case. Despite their effectiveness, encoder–decoder models may face reconstruction challenges, as compression into the bottleneck layer can result in the loss of crucial spatial features, limiting the recovery of fine-grained structures. 

To address this issue, our proposed architecture incorporates skip connections that directly transfer high-resolution features from the encoder to the decoder~\cite{UNET1}. This design preserves essential spatial correlations while simultaneously filtering out noise~\cite{GM2}. In addition to skip connections, our framework further enhances the architecture by aggregating feature maps from multiple hierarchical levels, thereby improving upon the standard model through multi-scale feature integration. This approach enables the network to capture richer contextual information, effectively combining fine-grained local details with broader structural patterns. The enhanced feature aggregation substantially improves estimation performance by preserving critical spatial structures and suppressing noise more effectively, even in scenarios with limited training data. Consequently, the proposed architecture offers a more robust and efficient solution for denoising tasks in large-scale systems.

\begin{figure}[!t]
        \vspace{-3mm}

    \centering
    \includegraphics[width=9 cm]{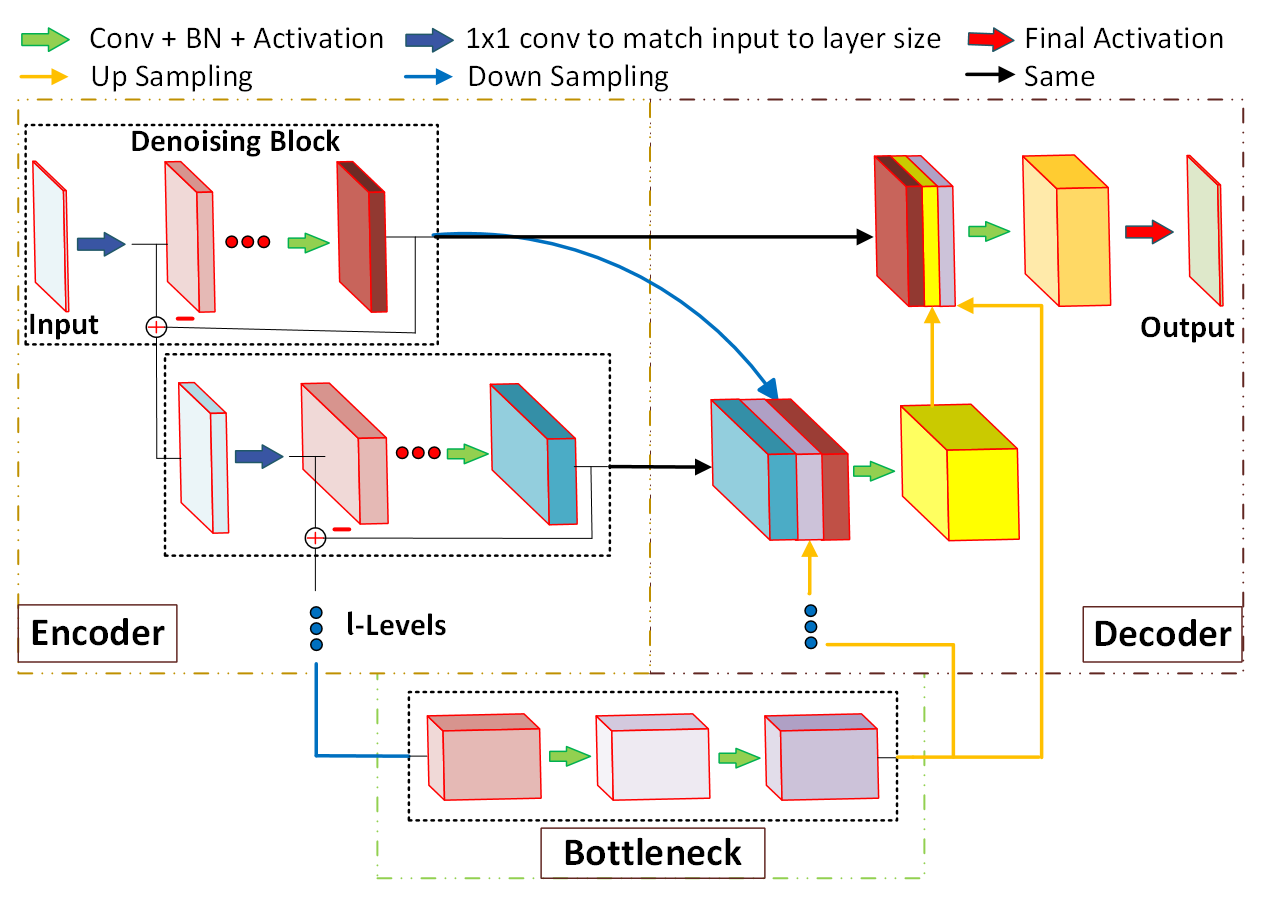}
    \caption{The proposed framework employs \(l\) levels, each with a dedicated denoising block in the encoder. Filters double, and the resolution halves at each level. Fused encoder and decoder features serve as input to each decoder layer. }
    \label{fig:4}
      \vspace{-3mm}

\end{figure}

\subsection{Proposed Model Architecture}
Our proposed architecture is designed to effectively capture the underlying signal features present in the dataset. The framework integrates a dedicated denoising module that employs a subtraction-based structure to enhance channel estimation by isolating the core signal components from additive noise. The architecture, illustrated in Fig. 3, adopts a multi-level encoder–decoder design to exploit spatial hierarchies to break the noise correlation, enabling more effective denoising and improved estimation accuracy. In the encoder, each level in the architecture consists of multi-layer Convolutional Neural Networks (CNNs) with distinct denoising blocks. The output of each higher-level block is downsampled by a factor of two and fed into the subsequent lower-level block, where the number of filters is doubled to capture increasingly complex features. We employ a MaxPooling layer with a (2,2) kernel in the encoder to downsample between levels. In the decoder, the output from lower-level blocks is upsampled to the current level, whereas the feature maps from the corresponding encoder blocks are downsampled to match that level. Additionally, the output from the same level in the encoder is concatenated with these other outputs, enabling effective multi-scale information fusion to preserve spatial structure. Moreover, we utilized bilinear interpolation for upsampling between levels in the network to smoothly increase the resolution of feature maps while maintaining spatial consistency. During the decoder phase, the filter size is halved, which helps to refine the reconstructed output. Finally, an activation function is applied to the aggregated features, adding nonlinearity and improving the model’s learning of complex mappings before the final output.

\subsection{Denoising Block Structure}
The CNN architecture is well-suited to extract underlying signal features from noisy data effectively. In \cite{DL1}, a subtraction-based mechanism was proposed to eliminate noise from the data, demonstrating effectiveness in denoising. However, this approach was restricted to learning models with identical dimensions, which is incompatible with the design of our proposed architecture. To address this limitation, the proposed denoising architecture introduces a \( 1 \times 1 \) convolutional layer at the beginning of each level in the encoder to align the input and output dimensions, thereby facilitating the application of the denoising mechanism. The denoising block architecture is composed of a CNN with \(i\) layers, where each layer consists of a convolution operation, batch normalization, and a non-linear activation function, as illustrated in Fig.~\ref{fig:4}. The denoising block applies a pointwise (\(1\times1\)) convolution to align the input with the network dimensions, followed by multiple convolutional layers. Finally, an element-wise subtraction is performed between the \(l\)-th level's block input \( \mathbf{I}_l \) and output \( \tilde{\mathbf{z}}_\theta(\mathbf{I}_l) \) to produce the denoised channel matrix, effectively preserving essential features. At the \(l\)-th level, it can be expressed as:

\begin{equation}
\mathbf{Z}_l = \mathbf{I}_l - \tilde{\mathbf{Z}}_\theta(\mathbf{I}_l),
\end{equation}
where \(\tilde{\mathbf{Z}}_\theta\) denotes the convolutional neural network and \(\theta\) represents its parameters. The primary challenge associated with cascaded channel estimation is the presence of Additive White Gaussian Noise (AWGN) \cite{OP1}. The simplest estimator for this task is the LS method, as it does not require statistical information. However, channel estimation using LS is often inadequate, as it is highly sensitive to noise, particularly in low Signal-to-Noise Ratio (SNR) conditions. Although it may still provide a rough estimate, its performance significantly degrades in noisy environments. Therefore, we use the LS estimator as input to our model, as it allows for a coarse estimate of the cascaded channel~\cite{DL1}. This initial estimate enables our model to focus on refining the channel representation, rather than learning it from scratch, thereby improving convergence and overall performance. Moreover, to accommodate the complex-valued nature of the data, each data point is decomposed into its real and imaginary components and provided as separate input channels to facilitate effective feature extraction.

\section{Spatial Correlation}

Spatial correlation refers to the degree of similarity between signals received or transmitted across different antenna elements, primarily influenced by their physical proximity and exposure to similar propagation paths~\cite{sp2}. When the RIS elements and BS antennas are closely spaced, the angular spread of the incident and reflected signals becomes narrow, leading to similar fading patterns across the arrays. Consequently, the cascaded channel experiences a certain level of spatial correlation~\cite{sp3}. This section provides a detailed discussion on the spatial correlation present in the cascaded channel and explores how it can be effectively leveraged to optimize the training process and reduce computational complexity.

\subsection{Spatial Correlation Between RIS Elements}
To model spatial correlation in the RIS elements, consider the channel between the \( m \)-th antenna at the BS and the \( N \) elements of the RIS, represented as a complex-valued vector
\(\mathbf{h}_m = [h_{m,1}, h_{m,2}, \dots, h_{m,N}] \in \mathbb{C}^N \). This channel vector can be further expressed as:

\begin{equation}
\label{eq_hm1}
    \mathbf{h}_m \sim \mathcal{N_C} \left( 0_N, [\mathbf{R}_b]_{m,m} \mathbf{R}_{r} \right),
\end{equation}
where \( \mathbf{R}_r \in \mathbb{C}^{N \times N} \) represent the correlation matrix of the RIS, which can be expressed as \(\mathbf{R}_r = \mathbb{E} \left\{ \mathbf{h}_m \mathbf{h}_m^H \right\} = \mathbb{E} \left\{ \mathbf{a}_r(\phi, \theta) \mathbf{a}_r(\phi, \theta)^H \right\}\) \cite{sp1}. To determine the correlation between elements, we select another RIS element \(n'\) and analyze the relationship between their respective channel responses, which can be expressed as:

\begin{align}
\mathbf{R}_{r_{[n,n']}} &= \frac{1}{{N M}} \mathbb{E} \Bigg[ 
e^{j \frac{2\pi}{\lambda}\big( \alpha_n \cos(\psi) \sin(\phi) d_H + \gamma_n \sin(\phi)d_V \big)}  \notag \\
&\quad \cdot \left( e^{j \frac{2\pi}{\lambda} \big( \alpha_{n'} \cos(\psi) \sin(\phi)d_H + \gamma_{n'} \sin(\phi)d_V \big)  }\right)^H
\Bigg].
\end{align}

Multiplying the exponential terms allows for further simplification of the above equation as follows:

\begin{align}
\!\!\mathbf{R}_{r_{[n,n']}} \!\!\!\!\!= \!\!\mathbb{E}\!\left[e^{j \frac{2\pi}{\lambda} \left[ (\alpha_{n}\! - \alpha_{n'}) \!\cos(\psi) \sin(\phi) d_H + (\gamma_n \!- \gamma_{n'}) \sin(\phi)d_V \right]} \right]\!.\!\!\!\!
\end{align}

We consider the case where \( \alpha_{n'}  = \alpha_{n} \), which means that the elements share the same row. This simplifies the expression for \( \mathbf{R}_{r_{[n,n']}} \), as the term that involves \( N_h \) disappears. The expectation can be reformulated using Equation (3) as:

\begin{equation}
\mathbf{R}_{r_{[n,n']}} = \int_{-\frac{\pi}{2}}^{\frac{\pi}{2}} \int_{-\frac{\pi}{2}}^{\frac{\pi}{2}} \frac{1}{\pi^2} \cdot e^{j \frac{2\pi}{\lambda}  ||\boldsymbol{d}_n -\boldsymbol{d}_{n'}|| \sin(\phi)  } d\phi \, d\psi.
\end{equation}

We now proceed to evaluate the integral of $\mathbf{R}_N$ with respect to the angular variables $\phi$ and $\psi$:

\begin{equation}
\mathbf{R}_{r_{[n,n']}} \!=\!\!\frac{1}{\pi}  \int_{-\frac{\pi}{2}}^{\frac{\pi}{2}} \!\! \, d\psi \cdot \! \left( \frac{1}{\pi} \int_{-\frac{\pi}{2}}^{\frac{\pi}{2}} \!\! e^{j \frac{2\pi}{\lambda}  ||\boldsymbol{d}_n -\boldsymbol{d}_{n'}|| \sin(\phi) } \, d\phi \!\! \right).
\end{equation}

When performing the integration over \(\phi\), the resulting expression is a classic oscillatory integral that evaluates to the Bessel function of the first kind, \(J_0\):

\begin{equation}
\mathbf{R}_{r_{[n,n']}} =  J_0 \left( { \frac{2\pi}{\lambda}  ||\boldsymbol{d}_n -\boldsymbol{d}_{n'}||  } \right) .
\end{equation}

The correlation function \(\mathbf{R}_{r_{[n,n']}}\) follows a zero-order Bessel function of the first kind \cite{sp3}. This expression captures the correlation between two RIS elements separated by \(||\boldsymbol{d}_n -\boldsymbol{d}_{n'}||\). The Bessel function shows that the correlation oscillates and gradually decays with increasing element separation, reflecting the spatial coherence properties of the scattered waves.

\subsection{Spatial Correlation Between RIS and BS}

In the previous subsection, we discussed the correlation between the \( m \)-th antenna and the RIS elements. However, the cascaded channel between the RIS and the BS, denoted as \( \mathbf{G}_k \in \mathbb{C}^{M \times N} \), consists of \( M \) antennas at the BS and \( N \) reflecting elements at the RIS. To determine the correlation between the \( m' \)-th antenna and the RIS elements, the corresponding channel coefficients can be expressed as: 

\begin{equation}
\label{eq_hm}
   \mathbf{h}_{m'} \sim \mathcal{N_C} \left( 0_N, [\mathbf{R}_b]_{m',m'} \mathbf{R}_{r} \right),
\end{equation}
Note that equation~\eqref{eq_hm} follows a structure similar to equation~\eqref{eq_hm1}, with the key distinction being in the variance terms \( [\mathbf{R}_b]_{m,m} \) and \( [\mathbf{R}_b]_{m',m'} \). These variance terms capture the spatial correlation among antennas, which are not independent but follow the correlation function \( [\mathbf{R}_b]_{m,m'} = \rho^{| m - m' |} \), where \( \rho \) represents the correlation coefficient. In conclusion, the channel matrix \( \mathbf{G}_i \in \mathbb{C}^{M\times N} \) exhibits spatial correlation, governed by the correlation structures of RIS and BS. Thus, in this paper, we leverage spatial correlation to minimize training complexity by proposing a patch-based training framework, which is discussed in detail in the following section.

\section{Patches based Training of Channel State Information}

Training machine learning models on edge devices is becoming increasingly popular, driven by the need for low-latency performance and enhanced data privacy \cite{FL}. Furthermore, it facilitates the development of robust and adaptive models that are tailored to local conditions, improving the overall performance of edge applications. However, training RIS-enabled XL-MIMO models on edge devices is challenging due to the size of the cascaded channel estimation matrices, \( \tilde{\mathbf{G}}^{\text{LS}}_{i} \in \mathbb{C}^{M \times N} \), where both \( M \) and \( N \) can range from hundreds to thousands \cite{XL3}. This results in massive data sizes, making storing and processing such large amounts of data on resource-constrained edge devices infeasible. To address this challenge, we propose a patch-based localized training mechanism. This method leverages spatial correlation between channels to reduce computational and memory demands, enabling feasible model training on edge devices despite their limited resources.

\subsection{Patches Based Training Framework}

The proposed patch-based training framework effectively addresses the challenges associated with training large models on resource-constrained devices. Rather than processing the entire high-resolution data point 
$\tilde{\mathbf{G}}^{\text{LS}}_{i} \in \mathbb{C}^{M \times N}$ at once, the framework partitions it into a few smaller, more manageable patches, as illustrated in Fig.~4. Training on these low-dimensional patches substantially reduces both memory and computational requirements, thereby enhancing overall training efficiency. This patch-based approach is particularly advantageous for edge devices, where computational and storage resources are limited. Moreover, by exploiting the spatial correlation inherent in the cascaded channel, neighboring patches exhibit strong correlation, as discussed in Section~IV. Therefore, it is not necessary to use all patches from a channel for training; a small number of representative patches per training example can provide sufficient information for effective model learning while reducing computational and memory requirements. These patches capture the essential features required for training, eliminating the need to process the full high-dimensional matrix $\tilde{\mathbf{G}}^{\text{LS}}_{i} \in \mathbb{C}^{M \times N}$.
This is particularly beneficial in the context of RIS-assisted XL-MIMO systems, where the number of RIS elements can range from hundreds to thousands. Our proposed approach significantly reduces computational complexity by processing only a small number of low-dimensional patches per sample, rather than the entire high-dimensional channel representation. This reduction in complexity enables edge devices to perform training at the edge, which would otherwise be infeasible when dealing with high-dimensional datasets.

As illustrated in Fig.~4, \(\tau\) represents the number of data points in \(\tilde{\mathbf{G}}^{\text{LS}}\). From each data point \( \tilde{\mathbf{G}}^{\text{LS}}_i \) in the training set, \(\mathcal{P}\) patches of size \(\mathcal{P}_x \times \mathcal{P}_y\) are extracted along with their corresponding labels \(\mathbf{G}_i\) to train the model, where \(\mathcal{P}_x\) and \(\mathcal{P}_y\) denote the patch dimensions along the \(x\)- and \(y\)-axes, respectively. We observe a trade-off between the number of patches \( \mathcal{P} \) extracted from each data point, their dimensions \( \mathcal{P}_x \) and \( \mathcal{P}_y \), and the available computational resources, including storage and processing capabilities. Nevertheless, increasing these parameters generally improves estimation accuracy, as elaborated in the simulation results section. However, they must be carefully chosen to satisfy the resource constraints of edge devices. For each patch, random starting coordinates \( x \) and \( y \) are selected within the valid range defined by the data dimensions \( M \) and \( N \), ensuring that the extracted patches do not overlap. This random but controlled selection strategy enhances the model's ability to learn spatially diverse features from different regions of the input data. After the patch generation process, a total of \( \mathcal{P}_{\text{Total}} = \mathcal{P} \times \text{(Total Training Data)} \) patches are generated and subsequently utilized for training. During the offline training phase, these patches and their corresponding labels are fed into the proposed architecture to optimize its parameters. Once the training is complete, the trained model is deployed for online inference. In contrast to the training stage, where the data are divided into smaller patches, the proposed architecture is designed to directly process the entire input channel matrix \( {\mathbf{G}}^{\text{LS}}_{k} \in \mathbb{C}^{M \times N} \) during inference, producing the estimated channel \( \tilde{\mathbf{G}}_{i} \in \mathbb{C}^{M \times N} \) without the need for patch extraction. This design choice significantly reduces inference time while maintaining high estimation accuracy.






\begin{figure}[!t]
    
    \vspace{-3mm}

    \centering
    \includegraphics[width=8 cm]{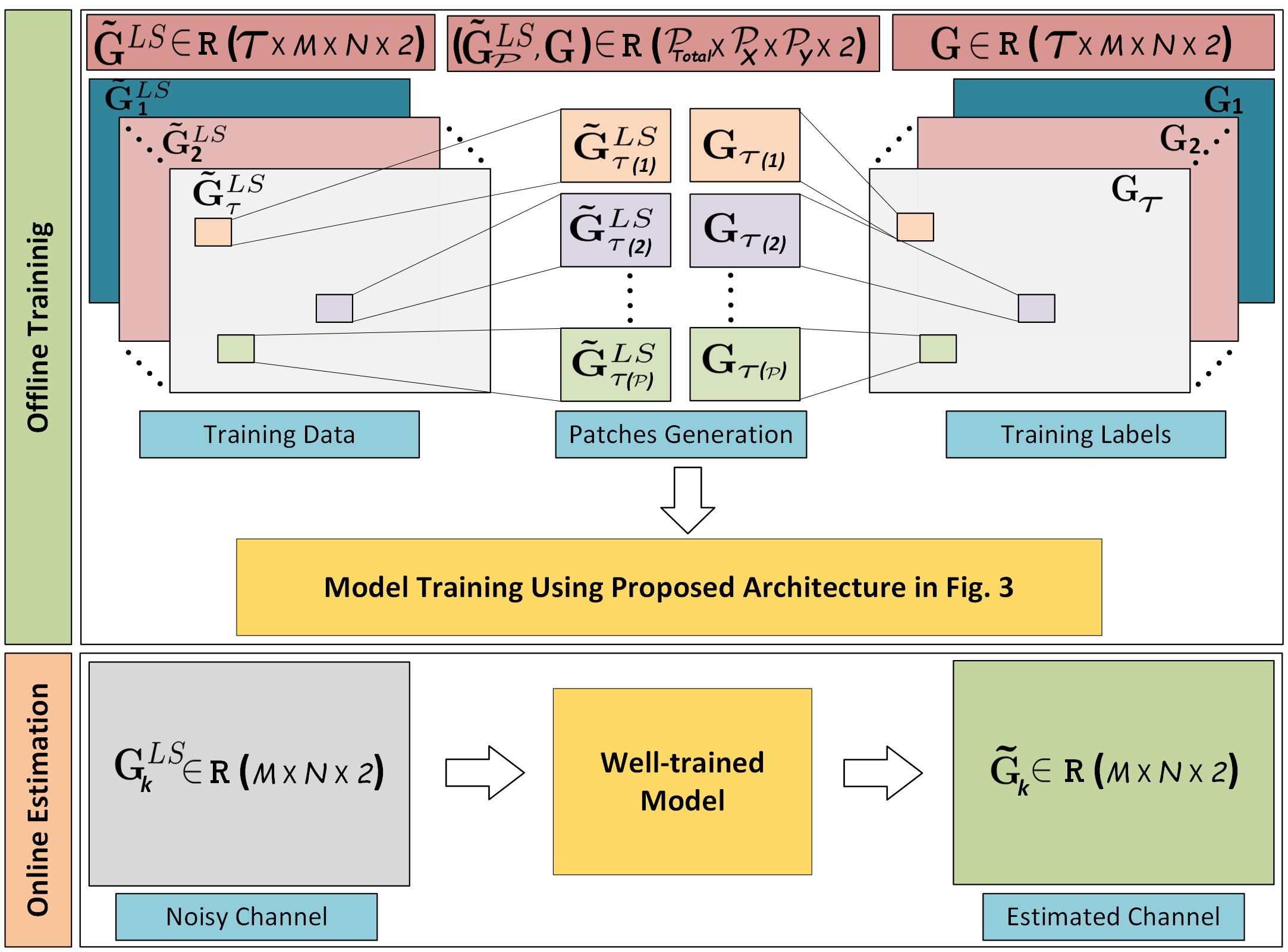}
    \caption{The training and inference stages of the proposed framework use matrices of dimension \(\mathbf{R} \), where the last dimension represents the real and imaginary parts of the channel.
 }
    \label{fig:4}
    
    \vspace{-3mm}

\end{figure}

\subsection{Efficacy of the Proposed Patch-Based Model Training}

Training the model using high-dimensional data points \( \tilde{\mathbf{G}}^{\text{LS}}_{i} \in \mathbb{C}^{M \times N} \) may not be the most efficient approach, given the inherent spatial correlation between antennas and RIS elements in the received signal at the BS, as discussed in the section III. From equations~(21) and~(27), it is evident that the signals received at the \( m \)-th and \( m' \)-th antennas of the BS exhibit a significant degree of correlation. Consequently, $\mathbf{h}_m$ and $\mathbf{h}_{m'}$, representing two different patches, exhibit similar spatial correlation characteristics. Therefore, training the deep learning model using $\mathbf{h}_m$ alone is sufficient to learn the underlying correlation structure shared by $\mathbf{h}_{m'}$, eliminating the need to explicitly include $\mathbf{h}_{m'}$ in the training process. In our proposed framework, the encoder compresses the input \( \mathbf{h}_m \) into a lower-dimensional latent space at the bottleneck, capturing the most important features and correlations in the data. This latent representation inherently encodes the spatial correlation between \( \mathbf{h}_m \) and \( \mathbf{h}_{m'} \). The decoder then reconstructs the output from this compact representation. However, since the spatial correlation is implicitly captured in the latent space, the decoder does not need direct access to \( \mathbf{h}_{m'} \) to understand its structure. The model can generalize and learn the dependencies, making it unnecessary to include both \( \mathbf{h}_m \) and \( \mathbf{h}_{m'} \) during training. This reduces model complexity while preserving its ability to learn the underlying correlation patterns of high-dimensional data points $\tilde{\mathbf{G}}^{\text{LS}}_{i} \in \mathbb{C}^{M \times N}$, ensuring effective performance in extremely large-scale RIS-based MIMO systems.
\section{Computational Complexity}
The computational complexity of the proposed method primarily depends on the encoder and decoder architectures. In the encoder, complexity is mainly driven by convolutional operations and downsampling, where each level applies two convolutions followed by max pooling. Given an input feature map of dimensions \( H \times W \times C^{\text{in}} \), a convolutional layer with \( C^{\text{out}} \) filters and a kernel size of \( K \times K \) incurs a computational complexity of \(\mathcal{O}(H W C^{\text{in}} C^{\text{out}} K^2) \). Since each level in the encoder of the proposed architecture comprises two convolutional layers, along with a \(1 \times 1\) convolution for dimensional alignment and a max-pooling operation, the total computational complexity at level~\(l\), denoted by \(\mathcal{O}_l^e\), is given by:

\begin{equation}
\mathcal{O}_l^e=\mathcal{O}(2 H_l W_l C^{\text{in}}_l C^{\text{out}}_l K^2 + 2H_l W_l C^{\text{out}}_l),
\end{equation}
Since \( 2H_l W_l C^{\text{out}}_l \) has negligible impact, it is omitted. The max-pooling operation halves the spatial dimensions at each level while doubling the feature channels, leading to the following feature map dimensions at level \( l \):

\begin{equation}
H_l = \frac{H_0}{2^l}, \quad W_l = \frac{W_0}{2^l}, \quad C^{in}_l = 2^l C_0^{out},
\end{equation}
where \( H_0 \times W_0 \) represents the spatial dimensions of the input feature map. By summing over all levels from \( l = 0 \) to \( L - 1 \), the total computational complexity of the encoder, represented as \( \mathcal{O}^e \), can be expressed as:

\begin{equation}
\mathcal{O}^e =\mathcal{O} \left( \sum_{l=0}^{L-1} 2 \frac{H_0 W_0}{4^l} C^{in}_l (2C^{in}_l) K^2 \right).
\end{equation}

Substituting \( C^{in}_l = 2^l C_0^{out} \), the complexity simplifies to:

\begin{align}
\mathcal{O}^e &= \mathcal{O} \left( 4 H_0 W_0 (C_0^{\text{out}})^2 K^2 \sum_{l=0}^{L-1} \frac{4^l}{4^l} \right) \notag \\
              &\approx \mathcal{O} \left( H_0 W_0 (C_0^{\text{out}})^2 K^2 L \right).
\label{e}
\end{align}

\begin{table}[t]
\centering
\caption{Computational Complexity of Different Estimation Algorithms}
\begin{tabular}{|c|c|c|c|}
\hline
\textbf{Algorithms}  & \textbf{CDRN}  & \textbf{SSL} & \textbf{Proposed}  \\ \hline
\textbf{FLOPs}  & \(4.54 \times 10^{12}\)       &  \(1.55 \times 10^{11}\)   &   \(6.73 \times 10^{7}\)             \\ \hline

\end{tabular}
\label{tab:system_parameters}
\end{table}

Eq.~\eqref{e} shows that the encoder's computational cost scales linearly with the spatial resolution \(H_0 \times W_0\) and quadratically with the number of output filters \(C_0^{\text{out}}\). Moreover, the complexity of the bottleneck layer adopts the same convolutional structure, denoted by \( \mathcal{O}^b \), is given by:

\begin{equation}
\mathcal{O}^b = \mathcal{O}\left(H_L W_L (C^{in}_l)^2 K^2\right) \approx {H_0W_0 (C_0^{\text{out}})^2 K^2}.
\end{equation}

The decoder’s computational complexity mirrors the encoder in reverse, with feature maps up/downsampled, concatenated, and convolved, while output channels decrease toward the final layer. The complexity of the \(l\)-th decoder layer \(\mathcal{O}_l^d\) is given by:

\begin{equation}
\mathcal{O}_l^d=\mathcal{O}(H_l W_l C_{\text{concat}} C^{in}_l K^2).
\end{equation}

In the decoder, each level processes feature maps of size \(H_l \times W_l\) with \(C^{\text{in}}_l\) channels. Concatenation aggregates features from multiple levels, increasing the channel dimension to \(C_{\text{concat}} = L \times C^{\text{in}}_l\). The total decoder complexity is then given by \(\mathcal{O}^d\):

\begin{equation}
\mathcal{O}^d=\mathcal{O} \left( \sum_{l=0}^{L-1} \frac{H_0 W_0}{4^l} L (C_{l}^{in})^2 K^2 \right),
\end{equation}
Substituting \( C^{in}_l = 2^l C_0^{out} \), the complexity simplifies to:

\begin{equation}
\mathcal{O}^d=\mathcal{O} \left( H_0 W_0 (C_0^{out})^2 K^2 L^2 \right).
\end{equation}

Overall, the computational complexity can be expressed as the sum of the complexities of the encoder, bottleneck, and decoder. Thus, the total complexity \(\mathcal{O}_{\text{total}}\) becomes:

\begin{align}
\label{ov}
\mathcal{O}_{\text{total}} = \ & \mathcal{O}\left(H_0 W_0 (C_0^{\text{out}})^2 K^2 L\right) 
+{H_0W_0 (C_0^{\text{out}})^2 K^2} \nonumber \\
& + \mathcal{O}\left(H_0 W_0 (C_0^{\text{out}})^2 K^2 L^2\right).
\end{align}

From \eqref{ov}, it is evident that the overall computational complexity scales directly with the size of the input feature map $H_0 \times W_0$, indicating that leveraging a patch-based training approach can substantially reduce the computational cost of the proposed model. Moreover, Table~I shows that the benchmark schemes CDRN~\cite{DL1} and SSL~\cite{DL4} require significantly more floating-point operations (FLOPs) per input sample during training compared to our proposed patch-based framework.

\begin{table}[t]
\centering
\caption{Simulation parameters}
\begin{tabular}{|c|c|c|c|}
\hline
\textbf{Parameter}  & \textbf{Value} & \textbf{Parameter} & \textbf{Value}\\ \hline
Antennas (M)        & 1024   & RIS Elements (N)     & 128          \\ \hline
\(M_h\) & 32     & \(M_v\)         & 32             \\ \hline
\(N_h\) & 16     & \(N_v\)         & 8             \\ \hline
\(\eta_r\) & 10     & \(\eta_b\)         & 10             \\ \hline
\(\rho\) & 0.8     & Carrier Frequency         & 7.8 Ghz             \\ \hline
Base Filters (BF) & 32   &     \(n_{th}\) Level Filters & \(BF \times 2^l\)     \\ \hline
Activation Function & {Relu}     & Last Activation         & Linear             \\ \hline
Learning Rate (LR) & 0.004     & LR Scheduler         & Exp. Decay \\ \hline
 Levels       & 3     & Optimizer         & Adam \\ \hline
 Batch Size       & 32     & Epochs         & 40 \\ \hline
\end{tabular}
\label{tab:system_parameters}
\end{table}

\section{Simulation }
This section evaluates the performance of the proposed method. The simulation setup consists of 1024 antennas arranged as \(M_h = 32\) and \(M_v = 32\) planar array, and 128 RIS elements arranged as \(N_h = 16\) and \(N_v = 8\) planar array. The model uses 32 base filters, which double with each depth level in the encoder and are halved progressively in the decoder. Moreover, ReLU activation functions are applied throughout, with a linear function at the output layer. Training begins with an initial learning rate of 0.004, managed by an exponential decay scheduler and optimized via the Adam algorithm. Network configurations include a depth of 3 levels, with 2 convolutional layers per level, a batch size of 32, and 40 training epochs.

Among existing studies, we identified CDRN~\cite{DL1} and SSL~\cite{DL4} as the most relevant works. However, these schemes are limited by their reliance on a relatively small number of antennas and RIS elements. Specifically, their training procedures involve processing full-dimensional matrices $\tilde{\mathbf{G}}^{(i)}_k \in \mathbb{C}^{M\times N}$ for each data point. Given the scale and complexity of XL-MIMO systems, where the number of antennas and RIS elements can reach hundreds or even thousands, training such models on large-scale datasets is computationally and memory intensive. This makes it challenging to train these models using full-dimensional channel matrices on existing computing systems. To address these limitations and ensure a fair evaluation, we trained the benchmark models using the same set of patches extracted from the dataset as those used in our model. This strategy aligns with edge device constraints while maintaining a consistent comparison with the proposed method. Moreover, several classical estimation techniques, including LS~\cite{LS}, LMMSE~\cite{LMMSE}, and B-LMMSE~\cite{BLMMSE}, are included as relevant benchmarks. The performance of the proposed framework is evaluated using the normalized mean squared error (NMSE), defined as \(\text{NMSE} = \frac{\mathbb{E}\!\left[\| \mathbf{G} - \hat{\mathbf{G}}_{\text{proposed}} \|_F^2\right]}{\mathbb{E}\!\left[\| \mathbf{G} \|_F^2\right]}\) where $\hat{\mathbf{G}}_{\text{proposed}}$ and $\mathbf{G}$ denote the estimated and ground-truth channels, respectively.

In this paper, we generated a synthetic dataset based on the system models described in Section III. The dataset was created with consideration for the memory constraints of edge devices. Specifically, a total of 10,000 data samples were generated, and from each sample, a representative data patch was extracted to encapsulate the underlying characteristics of the corresponding data point. As the antenna array has dimensions of \( 32 \times 32 \) and the RIS elements are arranged as \( 16 \times 8 \), the patch sizes were selected as \( 16 \times 32 \) and \( 32 \times 32 \) to align with the spatial correlation properties between the antenna and RIS elements. The dataset was split into training and validation sets, with 70\% allocated for training and 30\% for validation. Additionally, we created a separate testing dataset for different SNRs, with a sample size of 2,000 for each SNR value. Each point in the figures represent the average NMSE over 2,000 test samples.

\begin{figure}[!t]
    \vspace{-2mm}

    \centering
    \includegraphics[width=8 cm]{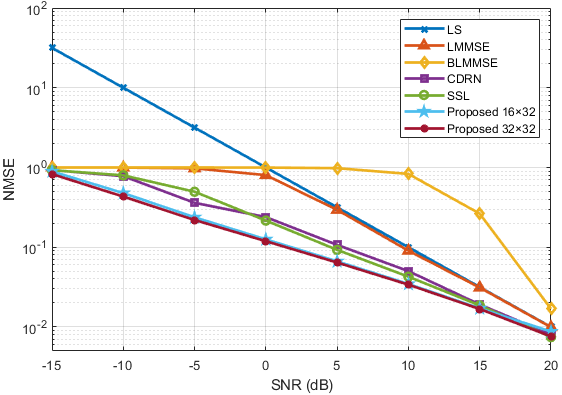}
    \caption{Performance comparison of the proposed scheme with benchmark methods and with varying patch sizes.
}
    \label{fig:nmse_comparison}
    \vspace{-5mm}
    
\end{figure}

Figure 5 presents the performance comparison between the proposed scheme and benchmark algorithms across an SNR range of -15 dB to 20 dB. At low SNR levels ($-15$~dB to $-10$~dB), where noise dominates, the proposed scheme outperforms existing methods, achieving an error reduction of approximately $13.65$~dB compared to LS, $0.8$~dB over LMMSE and BLMMSE, and $0.5$~dB over CDRN and SSL. As the SNR increases to $-5$~dB and $0$~dB, the proposed scheme consistently outperforms LS by up to $11.58$~dB, LMMSE and BLMMSE by $ 6.48$-$8.29$~dB, and learning-based methods CDRN and SSL by $3.02$ and $2.58$~dB, respectively, demonstrating superior generalization and feature extraction capabilities. In the moderate-to-high SNR region (5--20~dB), the performance advantage of the proposed model persists even as competing methods approach saturation. At 10~dB, it outperforms LS, LMMSE, and BLMMSE by $4.69$, $4.26$, and $13.89$~dB, respectively, and achieves gains of $1.7$ and $1.0$~dB over CDRN and SSL. At 15~dB and 20~dB, the proposed scheme continues to show improvements, achieving NMSE reductions of $2.74$ and $1.18$~dB over LS, $2.69$--$1.16$~dB over LMMSE, and $0.56$--$0.23$~dB and $0.45$--$0.10$~dB over CDRN and SSL, respectively. These results demonstrate that the proposed model maintains its robustness and accuracy even under high-SNR conditions. Additionally, we evaluate the proposed method using two patch sizes, $16 \times 32$ and $32 \times 32$, to assess the impact of patch size on the model’s performance. The results indicate that the NMSE varies across patch sizes, with the $32 \times 32$ configuration consistently achieving lower NMSE values than the $16 \times 32$ case, showing improvements ranging from $0.4$~dB to $0.08$~dB. This improvement arises from the larger patch size capturing stronger spatial correlations, thereby reducing estimation errors (especially at low SNRs), albeit at the cost of higher complexity.

 \begin{figure}[!t]
    \vspace{-3mm}

    \centering
    \includegraphics[width=8.5 cm]{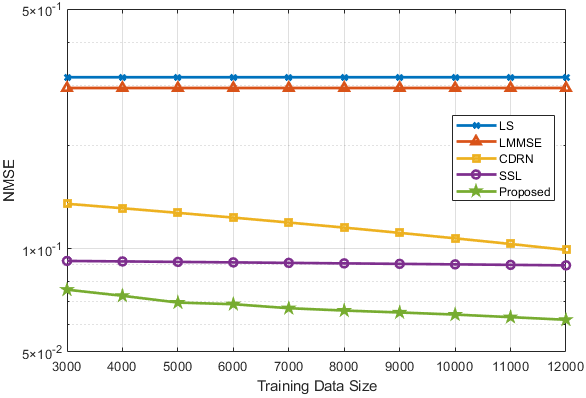}
    \caption{Compared performance by varying the dataset size used for model training.}
    \label{fig:training_comparison}
    \vspace{-3mm}

\end{figure}

Figure~6 compares the performance of the proposed framework with benchmark algorithms across training dataset sizes from 3,000 to 12,000 samples. Since LS and LMMSE are analytical estimators that operate independently of training data, their performance remains unchanged, exhibiting NMSE values approximately $6.19$ dB and $5.88$ dB higher than those of the proposed framework, respectively, across all dataset sizes. In contrast, the data-driven models, CDRN and SSL, demonstrate a clear dependence on dataset size. For smaller datasets (3,000–5,000 samples), CDRN shows a performance degradation of about $2.5$–$2.6$~dB relative to the proposed scheme, while SSL lags by approximately $0.8$–$1.18$~dB. As the dataset size increases (6,000–9,000 samples), both models show gradual improvements due to enhanced feature learning and greater data diversity. Specifically, CDRN’s NMSE decreases from $2.49$~dB to $2.32$~dB, and SSL’s NMSE decreases from $1.41$~dB to $1.32$~dB. However, their performance remains inferior to the proposed framework by more than $2$~dB and $1$~dB, respectively. For larger datasets (10,000–12,000 samples), CDRN and SSL achieve NMSEs of approximately $2.03$~dB and $1.58$~dB, respectively, yet these values remain about $2$~dB and $1.5$~dB higher than those of the proposed model. The proposed framework consistently achieves lower NMSE with increasing dataset size, effectively leveraging larger datasets, capturing complex spatial correlations, and maintaining superior generalization compared with CDRN and SSL. Overall, increasing the dataset size improves estimation accuracy but also raises computational complexity and memory requirements, which must be carefully considered when deploying the model on resource-constrained edge devices.

 \begin{figure}[!t]
    \vspace{-3mm}

    \centering
    \includegraphics[width=8.5 cm]{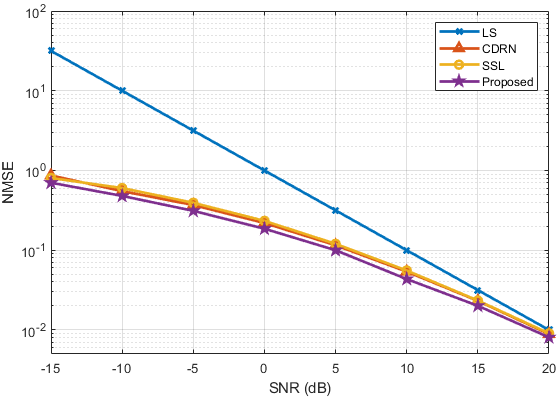}
    \caption{Performance comparison of the direct channel estimation from the user to BS.}
    \label{fig:direct_link_comparison}
    \vspace{-3mm}
    
\end{figure}

Figure~7 illustrates the performance comparison for direct channel estimation from the user to the BS across an SNR range of $-15$~dB to $20$~dB. The proposed framework consistently outperforms all benchmark algorithms across the entire SNR range. The LS estimator exhibits the poorest performance, with NMSE values approximately $16.35$~dB, $12.67$~dB, and $9.41$~dB higher than the proposed framework at $-15$~dB, $-10$~dB, and $-5$~dB, respectively. Even at moderate-to-high SNR levels ($0$–$20$~dB), LS remains significantly inferior, with the performance gap decreasing from $6.65$~dB at $0$~dB to $0.41$~dB at $20$~dB. In contrast, the learning-based models CDRN and SSL achieve closer results but still lag behind the proposed method. Specifically, at extremely low SNRs ($-15$~dB to $-10$~dB), CDRN performs $0.73$~dB and $0.8$~dB worse, while SSL trails by $0.41$~dB and $0.46$~dB, respectively. At $10$~dB, CDRN and SSL remain $0.3$~dB and $0.17$~dB below the proposed framework, while at $20$~dB, they nearly converge, with small gaps of $0.11$~dB and $0.1$~dB, respectively.
These results indicate that, while CDRN and SSL perform competitively at moderate-to-high SNRs, the proposed framework consistently achieves superior accuracy across all SNR levels, particularly under low-SNR and noise-dominant conditions, demonstrating strong generalization and noise resilience.

Figure~8 presents a comprehensive performance comparison of the proposed framework against benchmark schemes under both correlated and uncorrelated fading conditions for NLoS channels. The proposed models for both correlated and uncorrelated channels consistently outperform the corresponding CDRN and SSL benchmark schemes across all SNR levels. At low SNRs ($-15$~dB to $-10$~dB), the proposed correlated model achieves significantly lower error levels, outperforming CDRN and SSL under the same correlation by approximately $0.50$–$2.56$~dB and $0.49$–$2.67$~dB, respectively, while the uncorrelated configuration attains slightly higher NMSE improvements of $0.70$–$2.77$~dB and $0.46$–$2.79$~dB. As the SNR increases to moderate levels ($-5$~dB to $5$~dB), both correlated and uncorrelated versions of the proposed framework continue to demonstrate superior performance, with the correlated model achieving gains of approximately $2.15$–$2.21$~dB over CDRN and $3.55$–$1.58$~dB over SSL. These consistent improvements highlight the proposed model’s strong ability to capture complex spatial features and learn robust representations as signal reliability improves. At higher SNRs ($10$~dB to $20$~dB), where most models begin to saturate, the proposed correlated configuration maintains performance advantages of $1.69$–$0.23$~dB over CDRN and $1.07$–$0.10$~dB over SSL, while the uncorrelated framework achieves corresponding gains of approximately $1.15$–$0.28$~dB and $1.20$–$0.20$~dB. Overall, the correlated variants consistently outperform across all SNR regimes, demonstrating that spatial correlation enhances the model’s ability to exploit inter-element dependencies, leading to more accurate, stable, and noise-resilient channel estimation.

\begin{figure}[!t]
    \vspace{-3mm}
    \centering
    \includegraphics[width=8.5 cm]{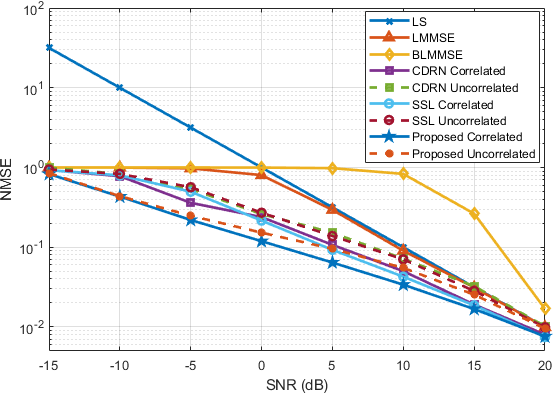}
    \caption{Comparative analysis between correlated and uncorrelated fading assumptions for NLoS channels, along with a comparison of results using different patch sizes for each case.}
    \label{fig:corr}
    \vspace{-3mm}

\end{figure}

Figure~9 illustrates the impact of increasing the number of phase shift matrices ($L$) on the NMSE performance, comparing the proposed scheme with LS, LMMSE, BLMMSE, CDRN, and SSL methods. As observed, increasing $L$ significantly improves estimation accuracy up to $L=128$, after which the performance gain becomes marginal. At $L=96$, the proposed method achieves an NMSE improvement of approximately $1.3$~dB over CDRN and $1.7$~dB over SSL, while maintaining a substantial margin of more than $4$~dB compared to the LS and LMMSE estimators. The most notable performance gain occurs at $L=128$, where the proposed scheme surpasses CDRN and SSL by around $2.2$~dB and $1.58$~dB, respectively. This improvement is attributed to better channel conditioning as $L$ approaches a full-rank LS matrix, enhancing the model’s learning capability. Beyond $L=128$, further increases in $L$ yield diminishing returns, as seen from the slower NMSE reduction. For higher values of $L$ ($144$, $160$, and $176$), the NMSE continues to decrease slightly, with the proposed method maintaining gains of approximately $1.2$–$1.5$~dB over CDRN and $1.2$–$1.4$~dB over SSL, indicating diminishing returns but consistently outperforming data-driven and analytical benchmarks.

\begin{figure}[!t]
    \vspace{-3mm}

    \centering
    \includegraphics[width=8.5 cm]{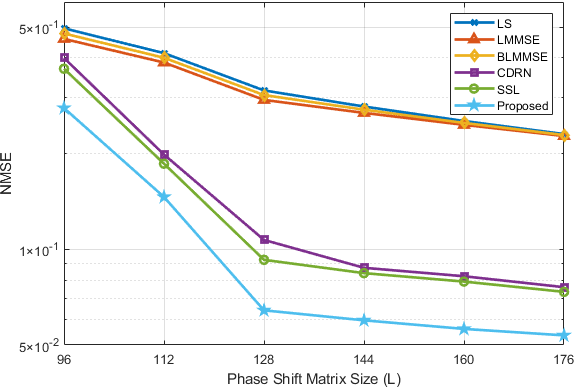}
    \caption{Performance comparison between LS and the proposed mechanism by varying the pilot overhead through changes in the phase shift matrix $L$ at an SNR of 5~dB.
}
    \label{fig:nmse_phase_shift}
    \vspace{-3mm}

\end{figure}

\section{ Conclusions }

In this work, we addressed the critical challenge of cascaded channel estimation in XL-MIMO systems empowered by RIS. Recognizing the limitations of traditional and existing deep learning approaches in resource-constrained environments, we proposed a novel, lightweight framework tailored for next-generation wireless networks. Using spatial correlation and patch-based training, our solution effectively reduces the dimensionality of the training data while preserving key features. We further enhance performance using our proposed architecture with integrated denoising capabilities, allowing accurate estimation even with limited training data. Extensive evaluations confirm that our framework enhances estimation accuracy and significantly reduces computational overhead, maintaining efficiency even as the number of antennas and RIS elements scales in XL-MIMO systems, making it well-suited for real-time deployment on edge devices. This pioneering approach lays the foundation for scalable, energy-efficient, and practical RIS-enabled XL-MIMO systems in 6G and beyond.

\bibliographystyle{IEEEtran}
\bibliography{ref}

@article{RIS6G,
  author    = {Kara, F. and Demir, \"O. T. and Bj\"ornson, E.},
  title     = {Can We Benefit from Reconfigurable Intelligent Surfaces in Upper Mid-Band 6G Networks? A Critical Look for Promising Use Cases},
  journal   = {arXiv preprint arXiv:2407.05754},
  year      = {2024}
}

@article{XL1,
  author    = {Wang, Z. and others},
  title     = {Extremely Large-Scale MIMO: Fundamentals, Challenges, Solutions, and Future Directions},
  journal   = {IEEE Wireless Communications},
  volume    = {31},
  number    = {3},
  pages     = {117--124},
  year      = {2024},
  month     = {June},
  doi       = {10.1109/MWC.132.2200443}
}

@article{XL2,
  author    = {Lei, H. and Zhang, J. and Wang, Z. and Ai, B. and Bjornson, E.},
  title     = {Near-Field User Localization and Channel Estimation for XL-MIMO Systems: Fundamentals, Recent Advances, and Outlooks},
  journal   = {IEEE Wireless Communications},
  year      = {2024},
  doi       = {10.1109/MWC.010.2400237}
}

@article{XL3,
  author    = {Bj\"ornson, E. and Kara, F. and Kolomvakis, N. and Kosasih, A. and Ramezani, P. and Salman, M. B.},
  title     = {Enabling 6G performance in the upper mid-band by transitioning from massive to gigantic MIMO},
  journal   = {arXiv preprint arXiv:2407.05630},
  year      = {2024},
  month     = {July}
}

@article{RIS,
  author    = {Bj\"ornson, E. and \"Ozdogan, \"O. and Larsson, E. G.},
  title     = {Reconfigurable Intelligent Surfaces: Three Myths and Two Critical Questions},
  journal   = {IEEE Communications Magazine},
  volume    = {58},
  number    = {12},
  pages     = {90--96},
  year      = {2020},
  month     = {December},
  doi       = {10.1109/MCOM.001.2000407}
}

@inproceedings{RIS1,
  author    = {Ahmed, S. and Abdelmawla, I. and Kamal, A. E. and Selim, M. Y.},
  title     = {Blockage Prediction for Mobile UE in RIS-assisted Wireless Networks: A Deep Learning Approach},
  booktitle = {MILCOM 2022 - 2022 IEEE Military Communications Conference (MILCOM)},
  year      = {2022},
  address   = {Rockville, MD, USA},
  pages     = {705--710},
  doi       = {10.1109/MILCOM55135.2022.10017894}
}

@article{OP1,
  author    = {Zhou, G. and Pan, C. and Ren, H. and Popovski, P. and Swindlehurst, A. L.},
  title     = {Channel Estimation for RIS-Aided Multiuser Millimeter-Wave Systems},
  journal   = {IEEE Transactions on Signal Processing},
  volume    = {70},
  pages     = {1478--1492},
  year      = {2022},
  doi       = {10.1109/TSP.2022.3158024}
}

@inproceedings{BLMMSE,
  author    = {Mishra, D. and Johansson, H.},
  title     = {Channel Estimation and Low-complexity Beamforming Design for Passive Intelligent Surface Assisted MISO Wireless Energy Transfer},
  booktitle = {ICASSP 2019 - 2019 IEEE International Conference on Acoustics, Speech and Signal Processing (ICASSP)},
  address   = {Brighton, UK},
  pages     = {4659--4663},
  year      = {2019},
  doi       = {10.1109/ICASSP.2019.8683663}
}

@article{OP2,
  author    = {Guo, H. and Lau, V. K. N.},
  title     = {Uplink Cascaded Channel Estimation for Intelligent Reflecting Surface Assisted Multiuser MISO Systems},
  journal   = {IEEE Transactions on Signal Processing},
  volume    = {70},
  pages     = {3964--3977},
  year      = {2022},
  doi       = {10.1109/TSP.2022.3193626}
}

@article{OP3,
  author    = {Chen, J. and Liang, Y.-C. and Cheng, H. V. and Yu, W.},
  title     = {Channel Estimation for Reconfigurable Intelligent Surface Aided Multi-User mmWave MIMO Systems},
  journal   = {IEEE Transactions on Wireless Communications},
  volume    = {22},
  number    = {10},
  pages     = {6853--6869},
  year      = {2023},
  month     = {October},
  doi       = {10.1109/TWC.2023.3246264}
}

@article{DL5,
  author    = {A. M. Elbir and A. Papazafeiropoulos and P. Kourtessis and S. Chatzinotas},
  title     = {Deep Channel Learning for Large Intelligent Surfaces Aided mm-Wave Massive MIMO Systems},
  journal   = {IEEE Wireless Commun. Lett.},
  volume    = {9},
  number    = {9},
  pages     = {1447--1451},
  year      = {2020},
  month     = sep,
  doi       = {10.1109/LWC.2020.2993699}
}

@article{DL1,
  author    = {C. Liu and X. Liu and D. W. K. Ng and J. Yuan},
  title     = {Deep Residual Learning for Channel Estimation in Intelligent Reflecting Surface-Assisted Multi-User Communications},
  journal   = {IEEE Trans. Wireless Commun.},
  volume    = {21},
  number    = {2},
  pages     = {898--912},
  year      = {2022},
  month     = feb,
  doi       = {10.1109/TWC.2021.3100148}
}

@article{DL2,
  author    = {M. H. Rahman and M. A. S. Sejan and M. A. Aziz and J.-I. Baik and D.-S. Kim and H.-K. Song},
  title     = {Deep Learning-Based Improved Cascaded Channel Estimation and Signal Detection for Reconfigurable Intelligent Surfaces-Assisted MU-MISO Systems},
  journal   = {IEEE Trans. Green Commun. Networking},
  volume    = {7},
  number    = {3},
  pages     = {1515--1527},
  year      = {2023},
  month     = sep,
  doi       = {10.1109/TGCN.2023.3237132}
}

@article{DL3,
  author    = {W. Shen and Z. Qin and A. Nallanathan},
  title     = {Deep Learning for Super-Resolution Channel Estimation in Reconfigurable Intelligent Surface Aided Systems},
  journal   = {IEEE Trans. Commun.},
  volume    = {71},
  number    = {3},
  pages     = {1491--1503},
  year      = {2023},
  month     = mar,
  doi       = {10.1109/TCOMM.2023.3239621}
}

@article{DL6,
  author    = {J. He and H. Wymeersch and M. Di Renzo and M. Juntti},
  title     = {Learning to Estimate RIS-Aided mmWave Channels},
  journal   = {IEEE Wireless Commun. Lett.},
  volume    = {11},
  number    = {4},
  pages     = {841--845},
  year      = {2022},
  month     = apr,
  doi       = {10.1109/LWC.2022.3147250}
}

@article{DL4,
  author    = {Z. Zhang and T. Ji and H. Shi and C. Li and Y. Huang and L. Yang},
  title     = {A Self-Supervised Learning-Based Channel Estimation for IRS-Aided Communication Without Ground Truth},
  journal   = {IEEE Trans. Wireless Commun.},
  volume    = {22},
  number    = {8},
  pages     = {5446--5460},
  year      = {2023},
  month     = aug,
  doi       = {10.1109/TWC.2023.3233970}
}

@article{LS,
  author    = {M. Biguesh and A. B. Gershman},
  title     = {Training-based MIMO channel estimation: A study of estimator tradeoffs and optimal training signals},
  journal   = {IEEE Trans. Signal Process.},
  volume    = {54},
  number    = {3},
  pages     = {884--893},
  year      = {2006},
  month     = mar
}

@book{LMMSE,
  author    = {S. M. Kay},
  title     = {Fundamentals of Statistical Signal Processing, Volume I: Estimation Theory},
  publisher = {Prentice-Hall},
  address   = {Upper Saddle River, NJ, USA},
  year      = {1993}
}

@article{FL,
  author    = {A. M. Elbir and S. Coleri},
  title     = {Federated Learning for Channel Estimation in Conventional and RIS-Assisted Massive MIMO},
  journal   = {IEEE Trans. Wireless Commun.},
  volume    = {21},
  number    = {6},
  pages     = {4255--4268},
  year      = {2022},
  month     = jun,
  doi       = {10.1109/TWC.2021.3128392}
}

@article{GM2,
  author    = {J. Zhao and Y. Wu and Q. Zhang and J. Liao},
  title     = {Two-Stage Channel Estimation for mmWave Massive MIMO Systems Based on ResNet-UNet},
  journal   = {IEEE Syst. J.},
  volume    = {17},
  number    = {3},
  pages     = {4291--4300},
  year      = {2023},
  month     = sep,
  doi       = {10.1109/JSYST.2023.3234048}
}

@article{AE1,
  author    = {G. E. Hinton and R. R. Salakhutdinov},
  title     = {Reducing the dimensionality of data with neural networks},
  journal   = {Science},
  volume    = {313},
  number    = {5786},
  pages     = {504--507},
  year      = {2006},
  month     = jul,
  doi       = {10.1126/science.1127647}
}

@inproceedings{UNET1,
  author    = {O. Ronneberger and P. Fischer and T. Brox},
  title     = {U-Net: Convolutional networks for biomedical image segmentation},
  booktitle = {Proc. Int. Conf. Med. Image Comput. Comput.-Assist. Intervent. (MICCAI)},
  address   = {Munich, Germany},
  pages     = {234--241},
  year      = {2015},
  month     = oct
}

@article{sp0,
  author    = {L. Sanguinetti and E. Björnson and J. Hoydis},
  title     = {Toward Massive MIMO 2.0: Understanding Spatial Correlation, Interference Suppression, and Pilot Contamination},
  journal   = {IEEE Trans. Commun.},
  volume    = {68},
  number    = {1},
  pages     = {232--257},
  year      = {2020},
  month     = jan,
  doi       = {10.1109/TCOMM.2019.2945792}
}

@article{sp1,
  author    = {W. -X. Long and M. Moretti and A. Abrardo and L. Sanguinetti and R. Chen},
  title     = {MMSE Design of RIS-Aided Communications With Spatially-Correlated Channels and Electromagnetic Interference},
  journal   = {IEEE Trans. Wireless Commun.},
  volume    = {23},
  number    = {11},
  pages     = {16992--17006},
  year      = {2024},
  month     = nov,
  doi       = {10.1109/TWC.2024.3449074}
}

@article{sp2,
  author    = {E. Björnson and L. Sanguinetti},
  title     = {Rayleigh Fading Modeling and Channel Hardening for Reconfigurable Intelligent Surfaces},
  journal   = {IEEE Wireless Commun. Lett.},
  volume    = {10},
  number    = {4},
  pages     = {830--834},
  year      = {2021},
  month     = apr,
  doi       = {10.1109/LWC.2020.3046107}
}

@article{sp3,
  author    = {G. -H. Li and D. -W. Yue and S. -N. Jin},
  title     = {Spatially Correlated Rayleigh Fading Characteristics of RIS-Aided mmWave MIMO Communications},
  journal   = {IEEE Commun. Lett.},
  volume    = {27},
  number    = {8},
  pages     = {2222--2226},
  year      = {2023},
  month     = aug,
  doi       = {10.1109/LCOMM.2023.3289959}
}

@article{zheng2024convolutional,
  author={Zheng, Peicong and Lyu, Xuantao and Wang, Ye and Gong, Yi},
  journal={IEEE Transactions on Wireless Communications}, 
  title={Convolutional Dictionary Learning-Based Hybrid-Field Channel Estimation for XL-RIS-Aided Massive MIMO Systems}, 
  year={2025},
  volume={24},
  number={11},
  pages={9085-9098},
  doi={10.1109/TWC.2025.3570631}}

@ARTICLE{lee2025nearfield,
  author={Lee, Jeongjae and Hong, Songnam},
  journal={IEEE Transactions on Wireless Communications}, 
  title={Near-Field LoS/NLoS Channel Estimation for RIS-Aided MU-MIMO Systems: Piece-Wise Low-Rank Approximation Approach}, 
  year={2025},
  volume={24},
  number={6},
  pages={4781-4792},
  doi={10.1109/TWC.2025.3544198}}

@ARTICLE{bian2026sparse,
  author={Bian, Xuechun and Xu, Wenbo and Wang, Yue and Yuen, Chau and Cai, Zhipeng},
  journal={IEEE Transactions on Communications}, 
  title={A Sparse-Based Channel Estimation Scheme for XL-RIS-Assisted Wireless Communications With Low Pilot Overheads}, 
  year={2026},
  volume={74},
  number={},
  pages={367-380},
  doi={10.1109/TCOMM.2025.3615781}}

@ARTICLE{zhong2025hybrid,
  author={Zhong, Jingbo and Jin, Yong and Yu, Mengqi and Peng, Xin and Lu, Zhanyong},
  journal={IEEE Communications Letters}, 
  title={Hybrid-Field Channel Estimation for RISs-Aided XL-MIMO Systems by Leveraging Signal Separation}, 
  year={2025},
  volume={29},
  number={11},
  pages={2626-2630},
  doi={10.1109/LCOMM.2025.3606124}}

\end{document}